\documentclass[american,aps,pre,twocolumn]{revtex4-1}
\usepackage[T1]{fontenc}
\usepackage[latin9]{inputenc}
\usepackage{color}
\usepackage{array}
\usepackage{verbatim}
\usepackage{textcomp}
\usepackage{amsmath}
\usepackage{amssymb}
\usepackage{graphicx}
\usepackage{esint}

\makeatletter

\providecommand{\tabularnewline}{\\}

\@ifundefined{textcolor}{}
{%
 \definecolor{BLACK}{gray}{0}
 \definecolor{WHITE}{gray}{1}
 \definecolor{RED}{rgb}{1,0,0}
 \definecolor{GREEN}{rgb}{0,1,0}
 \definecolor{BLUE}{rgb}{0,0,1}
 \definecolor{CYAN}{cmyk}{1,0,0,0}
 \definecolor{MAGENTA}{cmyk}{0,1,0,0}
 \definecolor{YELLOW}{cmyk}{0,0,1,0}
}

\@ifundefined{date}{}{\date{}}
\makeatother

\usepackage{babel}
\begin{document}

\title{Fluids confined in wedges and by edges: \\
virial series for the line-thermodynamic properties of hard spheres}

\author{Ignacio Urrutia$^{\dag}$}

\email{iurrutia@cnea.gov.ar}

\affiliation{$^{\dag}$Departamento de Física de la Materia Condensada, Centro
Atómico Constituyentes, CNEA, and CONICET. Av.Gral.~Paz 1499 (1650),
San Martín, Pcia.~de Buenos Aires, Argentina.}

\date{\today}
\begin{abstract}
This work is devoted to analyze the relation between the thermodynamic
properties of a confined fluid and the shape of its confining vessel.
Recently, new insights in this topic were found through the study
of cluster integrals for inhomogeneous fluids, that revealed the dependence
on the vessel shape of the low density behavior of the system. Here,
the statistical mechanics and thermodynamics of fluids confined in
wedges or by edges is revisited, focusing on their cluster integrals.
In particular, the well known hard sphere fluid, which was not studied
in this framework so far, is analyzed under confinement and its thermodynamic
properties are analytically studied up to order two in the density.
Furthermore, the analysis is extended to the confinement produced
by a corrugated wall. These results rely on the obtained analytic
expression for the second cluster integral of the confined hard sphere
system as a function of the opening dihedral angle $0<\beta<2\pi$.
It enables a unified approach to both wedges and edges. 
\end{abstract}
\maketitle

\section{Introduction\label{sec:Intro}}

The thermodynamic properties of fluids are influenced by the geometry
of either the vessel or substrate, which constrain the spatial region
where the molecules of the systems move. Several efforts are continuously
devoted to reach a detailed description of the response of fluids
to some of the simplest geometrical constraints, including the confinement
in pores with slit, cylindrical and spherical shapes, as well as the
case of fluids in contact with planar and curved walls. Even more,
the behavior of fluids adsorbed in wedges, at edges,\cite{Henderson_2002,Henderson_2004,Henderson_2004_b,Botan_2009}
and geometrically structured surfaces was the subject of several studies
during the last decade.\cite{Schneemilch_2003,Bryk_2003b,Schoen_2002}
The interest on confined inhomogeneous fluids covers a wide range
of complexity and particles size, which starts at the simplest one-atom
per molecule (e.g. the noble gases) and goes up to proteins, polymers
(including DNA molecules) and large colloids.\cite{Henderson_2006,Almenar_2011,Karl_2011,Lutsko_2012_b,Statt_2012}

The simplest model for the interaction potential between particles
is based on of hard spheres (HS), which reproduce the excluded volume
effect between particles. This framework was applied not only to the
study simple fluids, but also, to model the interaction between colloids.
The HS model is so significant that colloidal particles were synthesized
to mimic this interaction.\cite{Pusey_1986,Royall_2013} Furthermore,
recent simulations of confined HS have contributed with new interesting
insights to the glass transition.\cite{Mandal_2014} Even more, the
simplicity of HS make them suitable for theoretical development.\cite{Kedzierski_2011}
Thus, several approximate theories based on different kind of perturbative
approaches, for example mean field, density functional and different
type of series expansions, adopt HS as the reference system to study
more realist models of fluids.\cite{Hansen2006,Zhao_2012} Given its
simplicity, the HS system is particularly useful to elucidate the
relationship between the geometrical confinement of a fluid and its
thermodynamic properties.\cite{Urrutia_2014} Despite of its relevance
and simplicity, the exact or quasi-exact analytic results about the
HS homogeneous-fluids are really scare. In the last years, some analytic
expressions for the thermodynamic properties of geometrically confined
HS fluid were obtained by analyzing the shape dependence of the low-order
cluster integral. In this framework, it was studied the fluid confined
by spherical\cite{Urrutia_2012,Urrutia_2014,Urrutia_2014a}, cylindrical,
spheroidal and box-shaped, surfaces.\cite{Urrutia_2010b}

In this work I analyze the statistical mechanical and thermodynamic
properties of a fluid confined by edges and wedges on the basis of
the representation of its grand potential in powers of the activity.
In Sec. \ref{sec:ClusterinEdgeWedge} the edge/wedge confinement is
discussed and the functional dependence of the cluster integral on
the measures of the edge/wedge is described. The Sec. \ref{sec:WedgeThermo}
is dedicated to revisit the thermodynamics of the fluid in an edge/wedge
confinement, emphasizing the necessity of refer the properties of
the system to a particular choice of the reference region. This approach
is used in Sec. \ref{sec:Results} to analyze the HS system. Sec.
\ref{sec:SecondCluster} is devoted to obtain the analytic expression
of the angle-dependent second cluster integral for the HS fluid covering
the complete angular range which includes both edges and wedges. In
Sec. \ref{sec:Results} the resulting second cluster integral is used
to derive analytic expressions for the thermodynamic properties (pressure,
surface tension, line tension, surface- and linear- adsorptions) of
the confined HS fluid up to order two in density. The new expressions
for the line-tension and the line-adsorption show the dependence with
the opening dihedral angle. Finally, the low density behavior of the
HS fluid in contact with a corrugated wall is studied. The Sec. \ref{sec:End}
is devoted to the final remarks.

\section{Cluster integrals for a fluid confined by wedges\label{sec:ClusterinEdgeWedge}}

Let a system of HS particles with diameter $\sigma$ confined by a
hard external potential $\phi\left(\mathbf{r}\right)$ to either an
edge or wedge (dihedral) shape region $\mathcal{A}$. Trough current
work the open dihedron geometrical shape will be simply referred to
as dihedron. For this system at temperature $T$ it was recently
shown that the $i$-th cluster integral, $\tau_{i}$, takes the linear
form\cite{Hill1956,Urrutia_2014}
\begin{equation}
\tau_{i}/i!=b_{i}V-a_{i}A+c_{i}L\:.\label{eq:Taui}
\end{equation}
Here the dihedron $\mathcal{A}$ is characterized by its volume $V$,
its surface area $A$, the length of its edge $L$ and the opening
dihedral angle between faces, $\beta$ (inner to the fluid). The
volume coefficients $b_{i}$ are the well known Mayer's cluster integrals
for homogeneous systems and the area coefficients $a_{i}$ are related
to those introduced by Bellemans for a fluid adsorbed on an infinite
wall.\cite{Bellemans_1962,Bellemans_1962_b} The first cluster integral
is $\tau_{1}=b_{1}V$ with $b_{1}=1$, while the rest of coefficients
($i>1$) $b_{i}$ and $a_{i}$ are independent of $\beta$. For the
well known HS fluid, several $b_{i}$ and $a_{i}$ were evaluated
(for $i=2,3,4,5$ see Ref. \cite{Urrutia_2014}). Finally, regarding
the functions $c_{i}\left(\beta\right)$ they are still unknown even
at the lowest non-trivial order $i=2$.

\begin{figure}
\begin{centering}
\includegraphics[height=3.3cm]{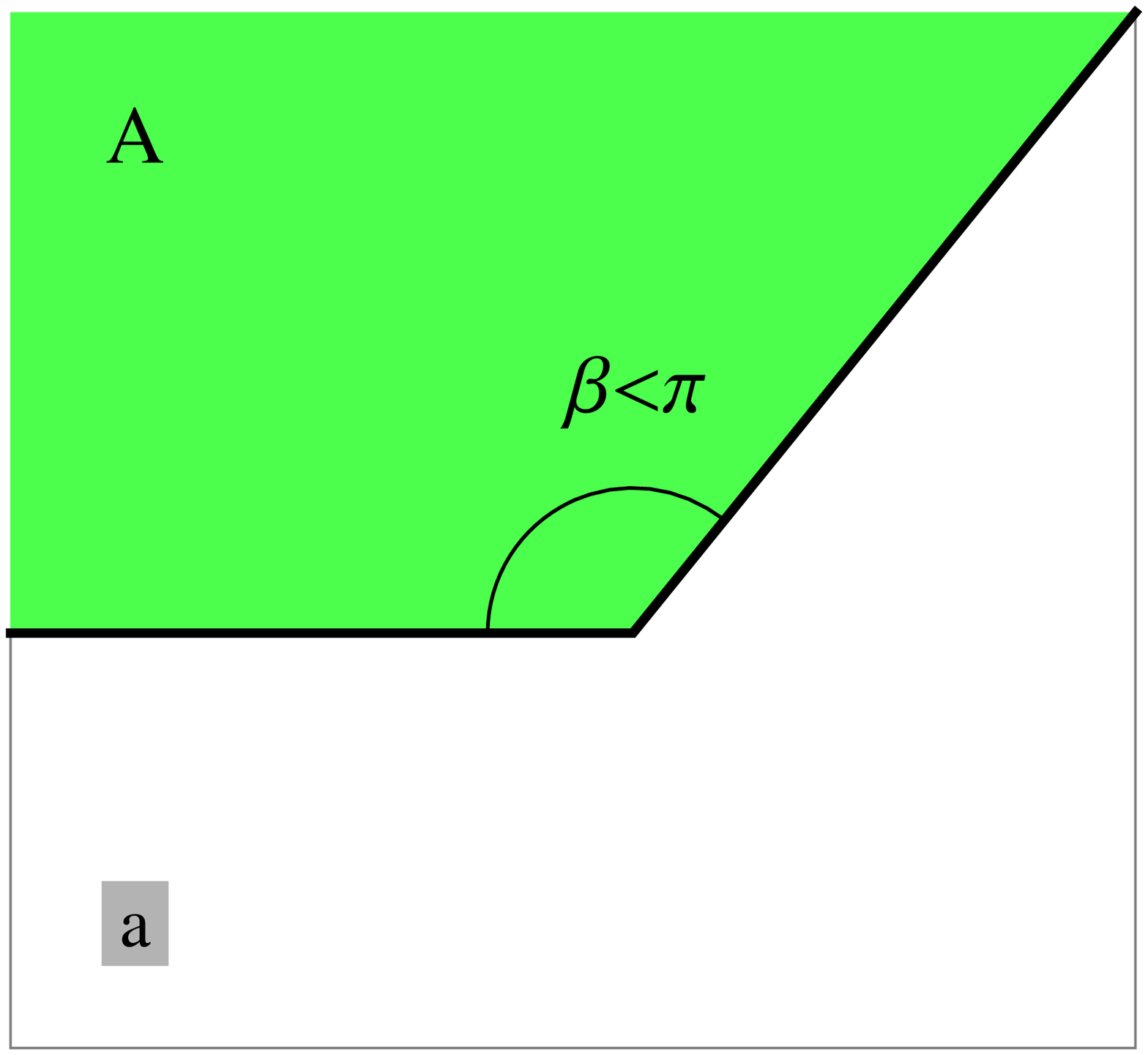}\includegraphics[height=3.3cm]{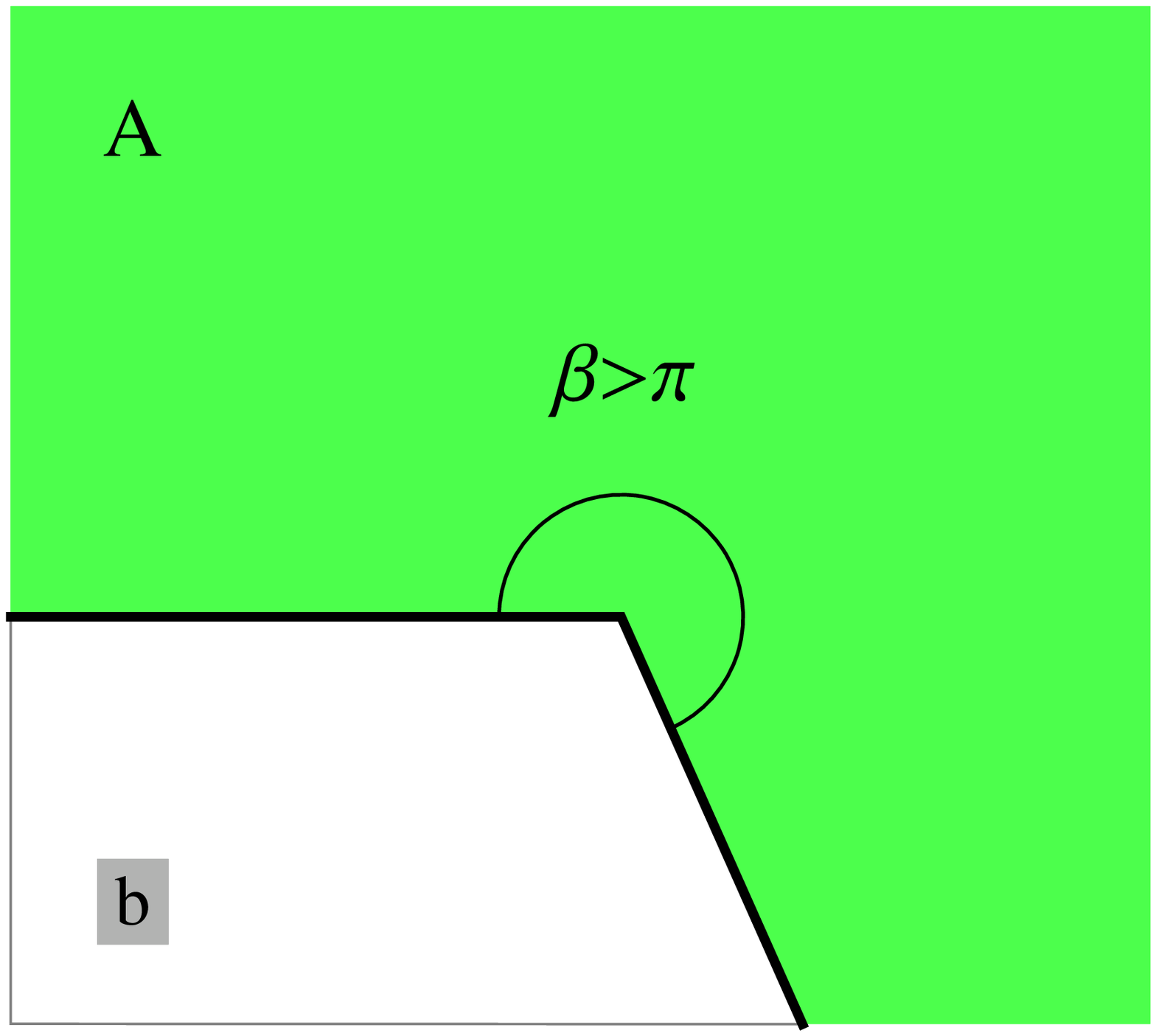}
\par\end{centering}

\protect\caption{Dihedral confinement for the particles. In the region $\mathcal{A}$
in light-gray (green) particles are free to move while the region
in white is forbidden. Note that no matter the value of the opening
angle both light-gray (green) and white regions have dihedral shape.
The dihedral shape of $\mathcal{A}$ determines that both Figs. a
and b are straight-edge type confinements.\label{fig:Adihedron}}
\end{figure}
In next sections two different types of edge/wedge confinement will
be analyzed. The dihedron $\mathcal{A}$, displayed in Fig. \ref{fig:Adihedron},
corresponds to the region available for the center of each particle.
This straight-edge confinement is defined by the Boltzmann factor
$\exp\left[-\phi\left(\mathbf{r}\right)/kT\right]=\Theta\bigl(\left|\mathbf{r}-\mathcal{C}\right|\bigr)$,
being $k$ the Boltzmann constant, $\Theta\bigl(x\bigr)$ the Heaviside
function {[}$\Theta\bigl(x\bigr)=1$ if $x>0$ and zero otherwise{]},
$\mathcal{C}=\mathcal{A}\setminus\mathbb{R}^{3}$ the complement dihedral
region, and $\left|\mathbf{r}-\mathcal{C}\right|$ the shortest distance
between $\mathbf{r}$ and $\mathcal{C}$. 
\begin{figure}
\begin{centering}
\includegraphics[height=4cm]{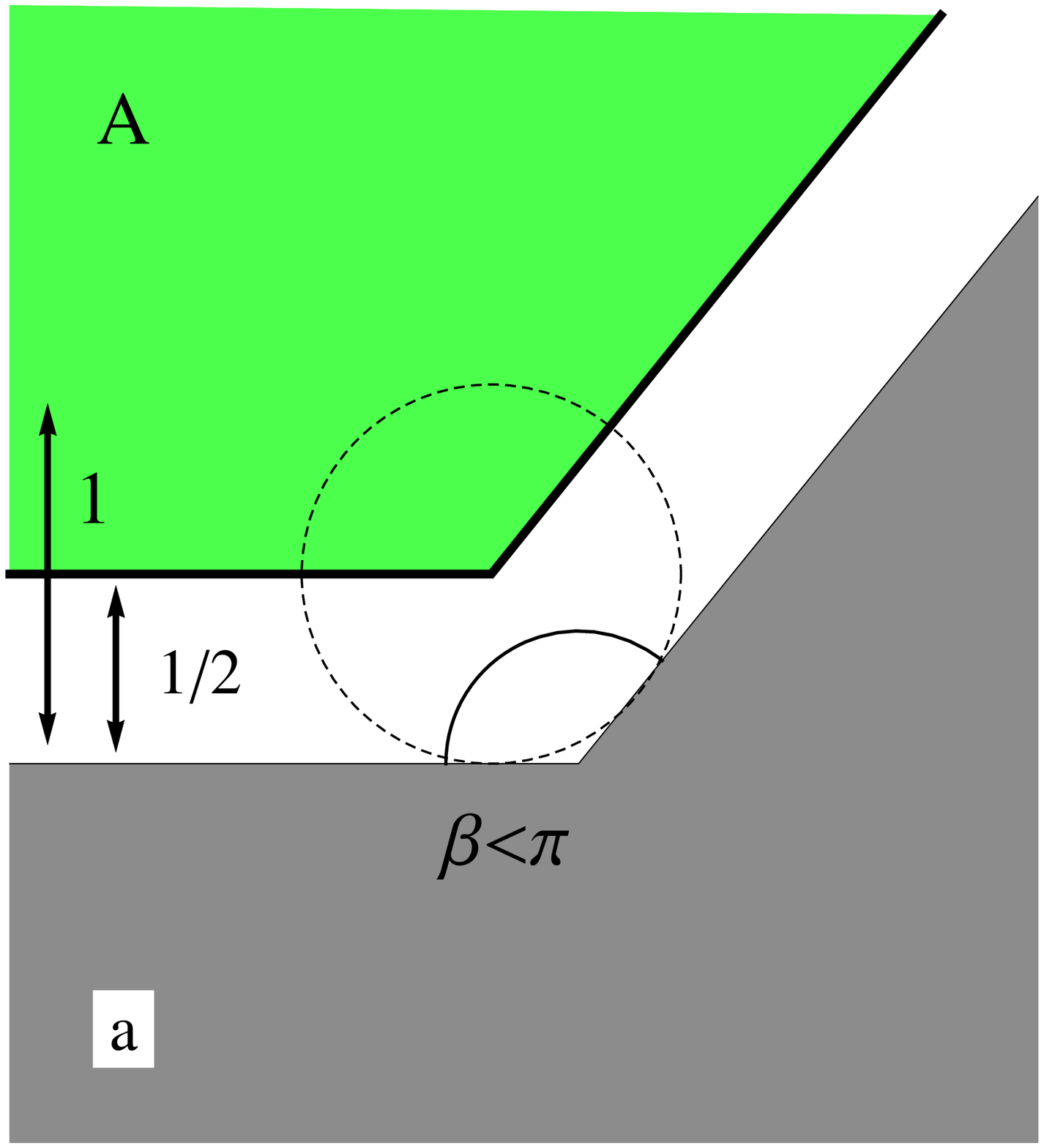}\includegraphics[height=4cm]{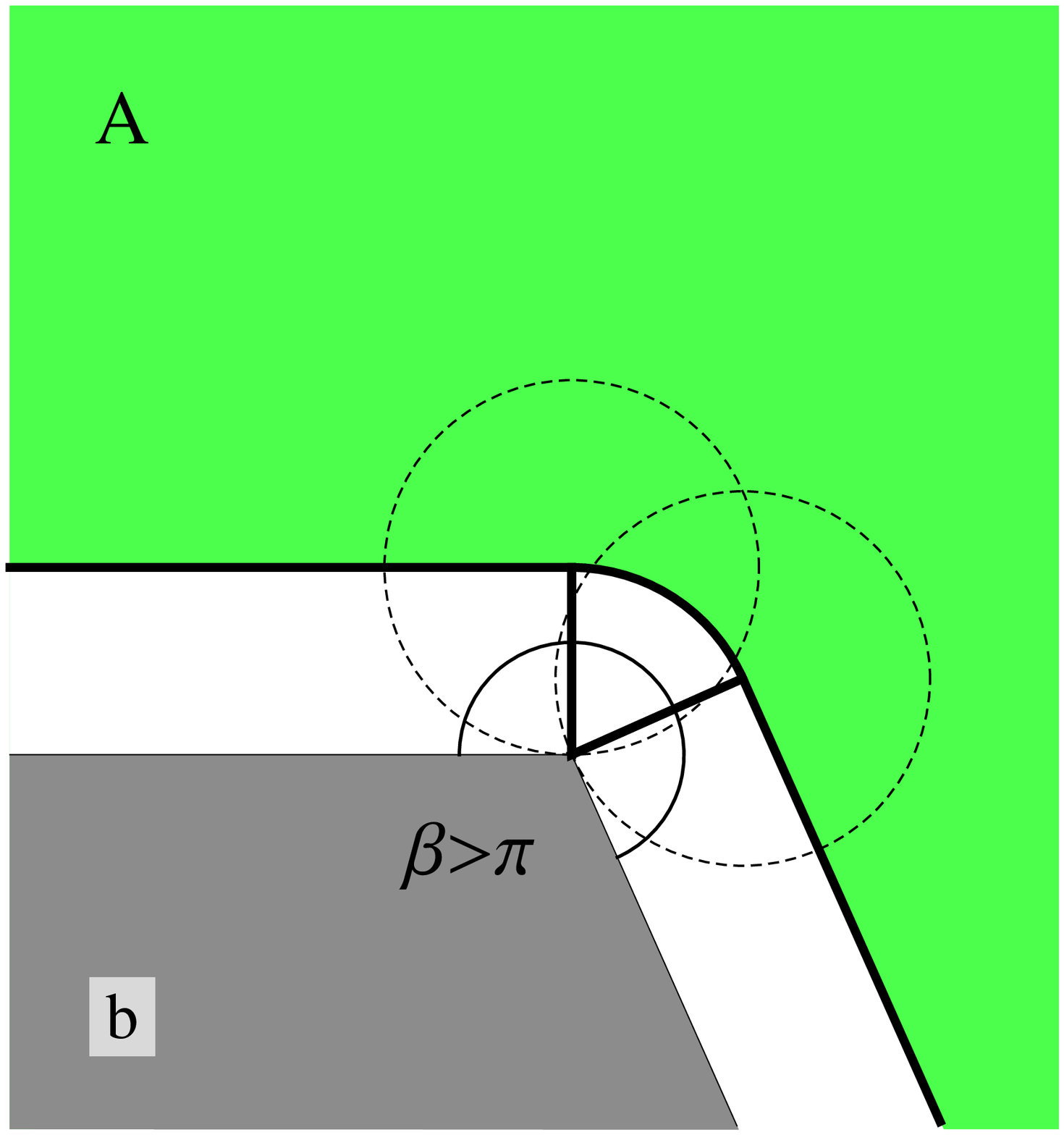}
\par\end{centering}

\protect\caption{Dihedral and quasi-dihedral confinement induced on the fluid by a
solid dihedron with solid-particle finite closest distance. Painted
with darker gray is the solid dihedron, lighter gray (green) corresponds
to the fluid and the empty region is in white. The draw in Fig. a
is for the case $\beta<\pi$ while Fig. b corresponds to the case
$\beta>\pi$. Dashed circles represent the hard repulsion wall-particle
distance for particles lying near the edge. The arrows show characteristic
lengths in $\sigma$ units. Note that $\mathcal{A}$ is a dihedron
in Fig. a, i.e. it a a straight-edge confinement, even, $\mathcal{A}$
in Fig. b corresponds to a rounded-edge confinement. \label{fig:HWallEdge}}
\end{figure}
A different type of confinement is defined by the Boltzmann factor
$\exp\left[-\phi\left(\mathbf{r}\right)/kT\right]=\Theta\left(\left|\mathbf{r}-\mathcal{C}\right|-\frac{\sigma}{2}\right)$
with $\mathcal{C}$ the solid dihedral region and $\frac{\sigma}{2}$
the minimum solid-particle distance. This last case is drawn in Fig.
\ref{fig:HWallEdge} where $\mathcal{C}$ is dark and the forbidden
region between $\mathcal{A}$ and $\mathcal{C}$ is shown in white.
From a comparison between Fig. \ref{fig:Adihedron} and Fig. \ref{fig:HWallEdge}
one notes that region $\mathcal{A}$ is a dihedron not only in Figs.
\ref{fig:Adihedron} a and b, but also in Fig. \ref{fig:HWallEdge}
a. Therefore, the case represented in Fig. \ref{fig:HWallEdge}a also
corresponds to a straight-edge confinement. On the contrary, the region
$\mathcal{A}$ shown in Fig. \ref{fig:HWallEdge}b is not a dihedron
because $\mathcal{A}$ has a curved end-of-fluid surface. This situation
corresponds to a rounded-edge confinement. Since Eq. (\ref{eq:Taui})
was originally derived for straight-edges, its extension to include
the rounded-edge case is presented in Sec. \ref{sub:CurvatureCorrection}.
Note that the above discussion regarding different type of edge/wedge
confinements essentially concerns to the confining potential but not
with the inter-particles potential.

\section{Thermodynamics of a fluid in a edge/wedge confinement\label{sec:WedgeThermo}}

To analyze the thermodynamic properties of the systems presented in
Figs. \ref{fig:Adihedron} and \ref{fig:HWallEdge} it is necessary
to adopt a reference region (RR) $\mathcal{R}$. It should be underlined
that for studying inhomogeneous fluids is crucial to clearly establish
the adopted $\mathcal{R}$, which fixes the position and shape of
the surface of tension, $\partial\mathcal{R}$. This issue is as fundamental
as to clearly establish the system of reference in the study of a
mechanical system. In this work it is adopted the density based RR
(d-RR), i.e. the region where the one-body density distribution is
non-null. This is in fact the region $\mathcal{A}$ shown in Figs.
\ref{fig:Adihedron} and \ref{fig:HWallEdge}.

Let us consider a fluid confined by a hard wall edge at fixed $T$
and chemical potential $\mu$. Its grand potential taken with regard
to the d-RR with measures $M=\left(V,A,L\right)$ is given by
\begin{equation}
\Omega=-PV+\gamma A+\mathcal{T}L\:.\label{eq:Omega}
\end{equation}
Here, $P$ is the pressure, $\gamma$ is the wall-fluid surface tension
(or excess surface free-energy), and $\mathcal{T}$ is the edge-fluid
line-tension (or excess line free-energy). Naturally, the Mayer series
of the grand potential of the system is given by 
\begin{equation}
\frac{\Omega}{kT}=-\sum_{i\geq1}\frac{\tau_{i}}{i!}z^{i}\,,\label{eq:Omegaz}
\end{equation}
where $z=\Lambda^{-3}\exp\!\left(\mu/kT\right)$ is the activity of
the fluid and $\Lambda$ the de Broglie\textasciiacute s length. It
is generally accepted that Eq. (3) is rigorous as long as condensation
is excluded (see p.131 of the book of Hill\cite{Hill1956}). Here
it is sufficient to assume that it converges for $0<z<R$ (for some
$R>0$). Since both Eqs. (\ref{eq:Taui}) and (\ref{eq:Omega}) are
linear in the measures $M$, therefore, Eq. (\ref{eq:Omegaz}) gives
the power series in $z$ representation of the intensive thermodynamic
properties $P$, $\gamma$ and $\mathcal{T}$. The mean number of
particles in the system is $N=-\frac{z}{kT}\,\frac{\partial\Omega}{\partial z}$,
and thus one obtain
\begin{eqnarray}
N & = & \rho V+\Gamma_{A}A+\Gamma_{L}L\,,\label{eq:N}\\
 & = & \sum_{i\geq1}i\frac{\tau_{i}}{i!}z^{i}\,,\label{eq:Nz}
\end{eqnarray}
here, $\Gamma_{A}$ is the excess adsorption per unit area (of the
boundary of $\mathcal{A}$) and $\Gamma_{L}$ is the excess adsorption
per unit length. Again, linear relations (\ref{eq:Taui}, \ref{eq:N})
and the power series in $z$ shown in (\ref{eq:Nz}) enable us to
obtain the power series of the densities $\rho(z)$, $\Gamma_{A}(z)$,
and $\Gamma_{L}(z)$. A linear decomposition, similar to that found
for $\Omega$ and $N$, is also obtained for the fluctuation $\sigma_{_{N}}^{\,2}\equiv\left\langle N^{2}\right\rangle -N^{2}=z\frac{\partial N}{\partial z}$.
This provides the power series in $z$ for each term in $\sigma_{_{N}}^{\,2}$
that scales with $V$, $A$ and $L$. Further, standard methods enable
to transform the power series in $z$ to power series in $\rho$.\cite{Hill1956}
Notably, from the Eqs. (\ref{eq:Omega}) to (\ref{eq:N}) and the
thermodynamic definition of $N$ {[}see expression above Eq. (\ref{eq:N}){]}
one can deduce the Gibbs adsorption equation for the surface adsorption
$\Gamma_{A}=-\partial\gamma/\partial\mu$ and also its analogous,
the Gibbs adsorption equation for the linear adsorption 
\begin{equation}
\Gamma_{L}=-\partial\mathcal{T}/\partial\mu\,.\label{eq:GibbsAdsL01}
\end{equation}

The confinement of the systems drawn in Fig. \ref{fig:Adihedron}
is purely characterized by the region $\mathcal{A}$ where the density
distribution could be non-null. For this type of confinement the unique
simple choice for RR is $\mathcal{A}$ itself. On the other hand,
even when the systems shown in Fig. \ref{fig:HWallEdge} can also
be analyzed under the same density-based $\mathcal{R}$ other RR could
be adopted. The relation between the thermodynamic properties obtained
under different choices of $\mathcal{R}$ will be studied in a forthcoming
paper. It is worthwhile to note that by adopting the density-based
$\mathcal{R}$, the systems depicted in Fig. \ref{fig:Adihedron}a
and Fig. \ref{fig:HWallEdge}a are identical, and thus, they have
identical properties.

\section{Second order cluster integral for HS confined by wedges\label{sec:SecondCluster}}

Here I focus on the second cluster integral for a straight-edge confinement.
In this first step I consider a system more general than the HS, in
which a pair of particles interact through the potential $\psi\left(r\right)$
with finite range $\xi$ {[}$\psi(r>\xi)=0${]}. For this case, the
Eq. (\ref{eq:Taui}) also holds.\cite{Urrutia_2013_prep} The Mayer's
function of the fluid confined in a region $\mathcal{A}$ is $f\left(r\right)=\exp\left[-\psi\left(r\right)/kT\right]-1$
while its second cluster integral reads\cite{Hill1956} 
\begin{equation}
\tau_{2}=\int_{\mathcal{A}}f\left(r_{12}\right)d\mathbf{r}_{1}d\mathbf{r}_{2}\:,\label{eq:Tau2}
\end{equation}
with $\mathbf{r}_{1}$, $\mathbf{r}_{2}$ the position of each particle
of the pair and $r_{12}$ the distance between them. Given that $\mathcal{A}$
has one edge with inner-dihedral angle $\beta$ (see Figs. \ref{fig:Adihedron}
and \ref{fig:HWallEdge}a), the coefficient $c_{2}\left(\beta\right)$
is given by\cite{Urrutia_2013_prep}
\begin{equation}
2c_{2}\left(\beta\right)=-2\int_{\mathcal{D}_{1}}g_{1}\left(x\right)d\mathbf{r}^{(2)}+\int_{\mathcal{D}_{2}}g_{2}\left(\mathbf{r}^{(2)}\right)d\mathbf{r}^{(2)}\:,\label{eq:c2eHS}
\end{equation}
where $\mathbf{r}^{(2)}$ is in the plane orthogonal to the edge
direction while $g_{1}(x)$ and $g_{2}(x)$ are partial integrals
of $f\left(r_{12}\right)$. For the case $0<\beta<\pi/2$, the integration
domains are
\begin{eqnarray}
D_{1} & = & I\cup II\cup III\:,\label{eq:D1}\\
D_{2} & = & I\cup II\:.\label{eq:D2}
\end{eqnarray}
\begin{figure}
\begin{centering}
\includegraphics[width=5cm]{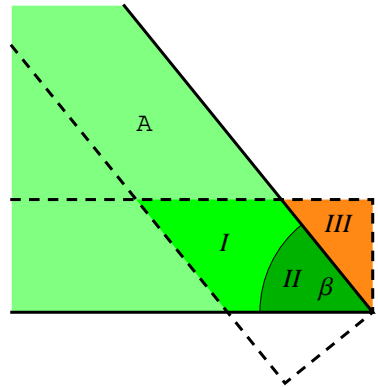}
\par\end{centering}

\protect\caption{Detail of the straight-edge of $\mathcal{A}$ with $0<\beta<\pi/2$.
In different gray-tones (green) are presented different regions near
the edge. The integration domains $\mathcal{D}_{1}$ and $\mathcal{D}_{2}$
of Eq. (\ref{eq:c2eHS}) are dissected in the regions $I$, $II$
and $III$. Region $III$ (orange) is outside of $\mathcal{A}$.\label{fig:IntegDom}}
\end{figure}
Regions\emph{ I}, \emph{II} and \emph{III} are shown in Fig. \ref{fig:IntegDom}.
The continuous line shows the boundary of the region $\mathcal{A}$,
$\partial\mathcal{A}$, which is defined by two planar faces. Parallel
to each face, at a distance $\xi$, there is a plane plot with dashed
line. A particle placed at $\mathbf{r}\in\mathcal{A}$ in the region
between dashed and continuous parallel lines is surrounded by an sphere
with radii $\xi$ that lies partially outside of $\mathcal{A}$. This
sphere separates the interacting region from the non-interacting region.

\begin{figure}
\begin{centering}
\includegraphics[width=5cm]{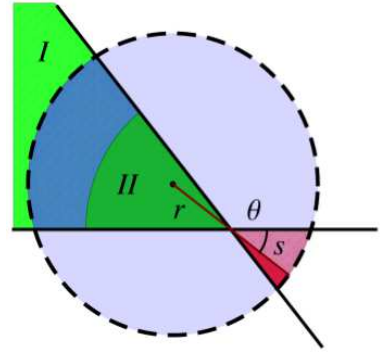}
\par\end{centering}

\protect\caption{\textcolor{black}{Exclusion sphere of radius $\sigma$ near the edge
of $\mathcal{A}$ for $\mathbf{r}$ in region $II$. The outer region
of the sphere with volume $v\left(r,\theta\right)$ is in lighter
gray (pink) while that with volume $v\left(r,\beta-\theta\right)$
is dark (red).\label{fig:vregion}}}
\end{figure}

From here on the analysis will only concern to the HS system and for
simplicity I fixed the hard repulsion distance $\sigma=\xi=1$. For
this system the volume and area coefficients of $\tau_{2}$ are: $b_{2}=-2\pi/3$
and $a_{2}=-\pi/8$. At present, the value of $c_{2}\left(\beta\right)$
is only known for two angles being $c_{2}(\pi/2)=-1/15$ and $c_{2}(3\pi/2)=-1/15$.\cite{Urrutia_2010b}
In addition, it is expected that $c_{2}\left(\beta\rightarrow\pi\right)=0$
due to the edge vanishes. Turning to Eq. (\ref{eq:Tau2}), for HS
$f\left(x\right)=-\Theta\left(1-x\right)$. By fixing $\mathbf{r}_{1}$,
$f(\bigl|\mathbf{r}_{1}-\mathbf{r}_{2}\bigr|)$ determines the so-called
exclusion unit sphere (centered at $\mathbf{r}_{1}$) for which $f\left(r_{12}\right)$
is non-null. Concerning to the functions $g_{1}\left(\mathbf{r}\right)$
and $g_{2}\left(\mathbf{r}\right)$, they measure the volume of the
portion of the exclusion unit sphere centered at $\mathbf{r}$ that
lies outside of $\mathcal{A}$ when $\mathbf{r}$ is near to $\partial\mathcal{A}$.
\begin{table}
\begin{centering}
\begin{tabular}{|c|c|c|c|c|}
\hline 
$j$ & $\:$ & $1$ & $2$ & $3$\tabularnewline
\hline 
f & $1$ & $2$ & $2$ & $2$\tabularnewline
\hline 
e & $0$ & $0$ & $0$ & $1$\tabularnewline
\hline 
 & \includegraphics[height=1cm]{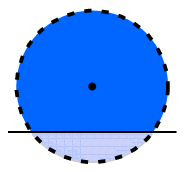} & \includegraphics[height=1cm]{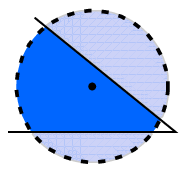} & \includegraphics[height=1cm]{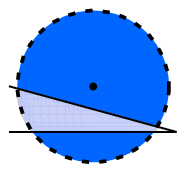} & \includegraphics[height=1cm]{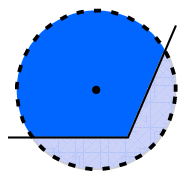}\tabularnewline
\hline 
\end{tabular}
\par\end{centering}

\protect\caption{Pictures that represent the exclusion sphere, in dashed line, for
different situations. The dark (blue) region is the overlap between
the sphere and $\mathcal{A}$ while in lighter gray (lighter blue)
is the region outside of $\mathcal{A}$.\label{tab:deFiguras}}
\end{table}
Table \ref{tab:deFiguras} summarizes the different shapes that the
outer portion of the exclusion sphere can take. There, f (e) is the
number of faces (edges) intersected by the unit sphere, $j$ is an
extra label for $\textrm{f}=2$, and for each $j$ corresponds a different
$g_{\textrm{f},j}\left(\mathbf{r}\right)$ function. Given $\mathbf{r}$
and a face of $\mathcal{A}$, I define the $x$ as the normal coordinate
of $\mathbf{r}$ relative to this face and positive in the inward
direction. Thus, $g_{1}\left(x\right)$ is the volume of that part
of the unit ball that lies in the semi-space with negative coordinate;
i.e., if $x>1$ then $g_{1}\left(x\right)=0$, if $-1\leq x\leq1$
then
\begin{equation}
g_{1}\left(x\right)=\frac{1}{3}\pi\left(2+x\right)\left(1-x\right)^{2}\:,\label{eq:g1HS}
\end{equation}
while for the case $x<-1$ $g_{1}\left(x\right)=4\pi/3$. On the other
side, given a position $\mathbf{r}$ and a pair of intersecting faces
that fix the coordinates $x$ and $x'$, $g_{2}\left(\mathbf{r}^{(2)}\right)$
is the volume of the unit ball that lies in the region where at least
one of these coordinates is negative. In the simple cases $g_{2}\left(\mathbf{r}^{(2)}\right)$
is equal to zero, $4\pi/3$, $g_{1}\left(x\right)$ or $g_{1}\left(x'\right)$.
For the less trivial cases and assuming that $x,x'>0$ one has
\begin{equation}
g_{2}(\mathbf{r}^{(2)})=\begin{cases}
g_{2,1}=g_{1}\left(x\right)+g_{1}\left(x'\right) & \textrm{if }j=1\:,\\
g_{2,2}=g_{2,1}-4\pi/3 & \textrm{if }j=2\:,\\
g_{2,3}=g_{2,1}-h\left(\mathbf{r}^{(2)}\right) & \textrm{if }j=3,\,\beta\leq\pi\,,\\
g_{2,3}=h\left(\mathbf{r}^{(2)}\right) & \textrm{if }j=3,\,\beta>\pi\,.
\end{cases}\:\label{eq:g2HS}
\end{equation}
Here, $h(\mathbf{r}^{(2)})$ is the intersecting volume between the
unit ball and the dihedron, that lies in the region where both coordinates
are negatives. To make further progress I analyze the case $0<\beta<\frac{\pi}{2}$
for which $h\left(\mathbf{r}^{(2)}\right)=v\left(r,\theta\right)+v\left(r,\beta-\theta\right)$,
with $r$ and $\theta$ the polar coordinates. An explicit expression
for $v$ was obtained by Rowlinson.\cite{Rowlinson_1963_c} After
taking into account trigonometric identities in Eq. (2.8) of Ref.\cite{Rowlinson_1963_c},
the following expression for $v$ is obtained
\begin{eqnarray}
 &  & v\left(r,\theta\right)=\text{ArcCot}\!\left[\frac{r\,\text{Cos}\left(\theta\right)}{\sqrt{1-r^{2}}}\right]\text{Sin}\left(\theta\right)\left(\frac{r^{3}}{3}\text{Sin}\left(\theta\right)^{2}-r\right)\!+\nonumber \\
 &  & \:\:\frac{2}{3}\text{ArcTan}\!\left[\sqrt{1-r^{2}}\text{Tan}\left(\theta\right)\right]\!+\!\frac{r^{2}}{6}\sqrt{1-r^{2}}\text{Sin}\left(2\theta\right)\,.\label{eq:vRowl}
\end{eqnarray}
Fig. \ref{fig:vregion} shows the exclusion sphere (dashed circle)
near the edge for the case $\textrm{f}=2$ and $j=3$ ($0<\beta<\frac{\pi}{2}$)
with $\mathbf{r}$ in the region $II$. There, the outer regions of
the sphere with volumes $v\left(r,\theta\right)$ and $v\left(r,\beta-\theta\right)$
are also presented. If $\mathbf{r}$ is in the region $II$ then $g_{2}=g_{2,3}$
(as is the case of Fig. \ref{fig:vregion}). On the other hand, if
$\mathbf{r}$ is in the region $I$ then $g_{2}=g_{2,1}$.
 Taking into account Eq. (\ref{eq:D2}), the integral of $g_{2}\left(\mathbf{r}^{(2)}\right)$
over $D_{2}$ reduces to $2\int_{I\cup II}g_{1}\left(x\right)d\mathbf{r}^{(2)}-2\int_{II}v\left(r,\theta\right)d\mathbf{r}^{(2)}$.
Using this expression and Eq. (\ref{eq:D1}) to transform Eq. (\ref{eq:c2eHS})
one found
\begin{equation}
c_{2}\left(\beta\right)=-\int_{III}g_{1}\left(x\right)d\mathbf{r}^{(2)}-\int_{II}v\left(r,\theta\right)d\mathbf{r}^{(2)}\:.\label{eq:c2eHS1}
\end{equation}
The integral over region $III$ gives $\pi\cot\left(\beta\right)/15$.
On the other hand, to solve the integral over region $II$ one introduces
the change of integration variable $r\rightarrow\sqrt{1-s^{2}}$ to
obtain $\left(1-\beta\cot\beta\right)/15$. Finally one found (for
$0<\beta\leq\pi/2$)
\begin{equation}
c_{2}\left(\beta\right)=-\frac{1}{15}\left[1+\left(\pi-\beta\right)\cot\beta\right]\:.\label{eq:c2eHS2}
\end{equation}

In principle, the procedure can be reproduced for each of the ranges
$\pi/2<\beta\leq\pi$, $\pi<\beta\leq3\pi/2$, and $3\pi/2<\beta<2\pi$,
but the expressions for $v$ become more complex, the domains of integration
change, and the integral of $v$ becomes harder to solve. Thus, an
alternative approach to analyze the range $\pi/2<\beta<2\pi$ is presented
below (see Sec. \ref{sub:OtherBetaRange}). Here I advance the result:
the Eq. (\ref{eq:c2eHS2}) applies to the complete range $0<\beta<2\pi$.

\subsection{Other ranges of beta\label{sub:OtherBetaRange}}

With the aim of circumvent the integral on $v$, I will use a statistical-based
analysis. Let us consider a region U of the space, composed by the
disjoint union of regions A and B. If one drops a particle randomly
in U the probability of finding it in A ($P_{\textrm{A}}$) or in
B ($P_{\textrm{B}}$) relate by 
\begin{equation}
1=P_{\textrm{A}}+P_{\textrm{B}}\:.\label{eq:PAPB}
\end{equation}
Let us consider a pair of (distinguishable) particles labeled as particle-$1$
and particle-$2$, randomly dropped in U. I will focus on the case
where particles form a cluster (i.e., they lie at a distance $r<1$)
and introduce $P_{\textrm{XY}}$ which is the probability that particle-$1$
lies in the region X while the particle-$2$ lies in the region Y.
Thus, for the case of a pair of clustered particles one found the
following relations
\begin{eqnarray}
1 & = & P_{\textrm{AA}}+P_{\textrm{BB}}+2P_{\textrm{AB}}\:,\label{eq:ProbAB}\\
P_{\textrm{AA}} & = & P_{\textrm{AU}}-P_{\textrm{AB}}\:,\label{eq:ProbAS}\\
P_{\textrm{BB}} & = & P_{\textrm{BU}}-P_{\textrm{BA}}\:,\label{eq:ProbBS}
\end{eqnarray}
with $P_{\textrm{AB}}=P_{\textrm{BA}}$. Each of these probabilities
is related to the volume of the coordinates phase space that corresponds
to a cluster integral by the expression 
\begin{equation}
P_{\textrm{XY}}=\tau_{\textrm{XY}}/\tau_{\textrm{UU}}\:.\label{eq:ProbXY}
\end{equation}
Here $\tau_{\textrm{XY}}$ is the second cluster integral with the
restriction that particle-$1$ is in $\textrm{X}$ while particle-$2$
is in $\textrm{Y}$ (with $\tau_{\textrm{XY}}=\tau_{\textrm{YX}}$).
Thus, any of the Eqs. (\ref{eq:ProbAB}) to (\ref{eq:ProbBS}) can
be translated to a relation between cluster integrals.

\begin{figure}
\centering{}\includegraphics[width=7cm]{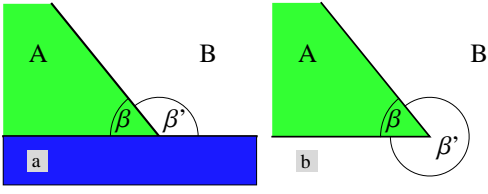}\protect\caption{Picture showing the statistical-based approach to relate the edge
term of $\tau_{2}\left(\mathcal{A}\right)$ with that of $\tau_{2}\left(\mathcal{B}\right)$.
Dihedral region $\mathcal{B}$, with opening angle $\beta'$, is in
white. For the region $\mathcal{A}$ (green) with $\beta<\pi/2$,
the term $c_{2}\left(\beta\right)$ is given in Eq. (\ref{eq:c2eHS2}).
Subfigure a is for the case of supplementary angles $\beta+\beta'=\pi$
with $\frac{\pi}{2}<\beta'<\pi$, while Subfigure b is for the case
of add-full angles $\beta+\beta'=2\pi$ with $\pi<\beta'<2\pi$.\label{fig:edgeforAB}}
\end{figure}

Let the region U be the semi-space $\textrm{S}$ that is partitioned
in two wedge-shaped regions labelled $\textrm{A}$ and $\textrm{B}$
with dihedral angles $\beta$ and $\beta'=\pi-\beta$ (with $0<\beta<\frac{\pi}{2}$),
respectively, as it is indicated in Fig. \ref{fig:edgeforAB}a. 
Based on Eqs. (\ref{eq:ProbAB}) and (\ref{eq:ProbXY}), the cluster
integral $\tau_{2}$ for the system constrained to different regions
are related one to each other. Thus, one finds
\begin{equation}
\tau_{\textrm{SS}}=\tau_{\textrm{AA}}+\tau_{\textrm{BB}}+2\tau_{\textrm{AB}}\:,\label{eq:T2Sparted}
\end{equation}
and from Eq. (\ref{eq:Taui})
\begin{eqnarray}
\tau_{\textrm{SS}} & = & 2b_{2}V_{\textrm{S}}-2a_{2}A\:,\label{eq:T2s-SS}\\
\tau_{\textrm{AA}} & = & 2b_{2}V_{\textrm{S}}\frac{\beta}{\pi}-2a_{2}A+2c_{2}\left(\beta\right)L\:,\label{eq:T2s-AA}\\
\tau_{\textrm{BB}} & = & 2b_{2}V_{\textrm{S}}\frac{\beta'}{\pi}-2a_{2}A+2c_{2}\left(\beta'\right)L\:,\label{eq:T2s-BB}
\end{eqnarray}
where $V_{\textrm{S}}$ is the volume of the semi-space, $A$ is the
area of the infinite plane, and $L$ is the length of the straight
line. Eqs. (\ref{eq:T2Sparted}, \ref{eq:T2s-SS}, \ref{eq:T2s-AA})
and (\ref{eq:T2s-BB}) imply that $\tau_{\textrm{AB}}$ should be
a linear function in $V_{\textrm{S}}$, $A$ and $L$. In fact, by
equating term by term in Eq. (\ref{eq:T2Sparted}) it is obtained
\begin{equation}
\tau_{\textrm{AB}}=-a_{2}A+2c_{mix}\left(\beta\right)L\:,\label{eq:T2s-AB}
\end{equation}
with $c_{mix}\left(\beta\right)=\frac{1}{2}\left[c_{2}\left(\beta\right)+c_{2}\left(\beta'\right)\right]$
which is \emph{symmetric} in the angles, i.e. $c_{2mix}\left(\beta\right)=c_{2mix}\left(\beta'\right)$.
On the same basis, Eqs. (\ref{eq:ProbAS}) and (\ref{eq:ProbBS})
are equivalent to
\begin{eqnarray}
\tau_{\textrm{AA}} & = & \tau_{\textrm{AS}}-\tau_{\textrm{AB}}\:,\label{eq:T2s-AA2}\\
\tau_{\textrm{BB}} & = & \tau_{\textrm{BS}}-\tau_{\textrm{AB}}\:,\label{eq:T2s-BB2}
\end{eqnarray}
which shows that $\tau_{\textrm{AS}}$ and $\tau_{\textrm{BS}}$ are
both linear in $V_{\textrm{S}}$, $A$ and $L$. Moreover, by combining
the Eqs. (\ref{eq:T2s-AA2}, \ref{eq:T2s-BB2}) and (\ref{eq:T2Sparted})
one found
\begin{equation}
\tau_{\textrm{SS}}=\tau_{\textrm{AS}}+\tau_{\textrm{BS}}\:.\label{eq:T2Sparted2}
\end{equation}
Finally, one introduces $2c_{aux}\left(\beta\right)$ and $2c_{aux}\left(\beta'\right)$
as the $L$ coefficients in $\tau_{\textrm{AS}}$ and $\tau_{\textrm{BS}}$,
respectively. Therefore, Eq. (\ref{eq:T2Sparted2}) shows that $c_{aux}$
is \emph{antisymmetric} in the angles, i.e. $c_{aux}\left(\beta'\right)=-c_{aux}\left(\beta\right)$.
It is clear that if $c_{aux}\left(\beta\right)$ is known one can
obtain $c_{aux}\left(\beta'\right)$, $c_{mix}\left(\beta\right)$
and also $c_{2}\left(\beta'\right)$. An interesting point is that
$\tau_{\textrm{AS}}$ {[}related to $c_{aux}\left(\beta\right)${]}
is simpler to solve than any of the integrals $\tau_{\textrm{BB}}$,
$\tau_{\textrm{AB}}$ or $\tau_{\textrm{BS}}$ {[}related to $c_{2}\left(\beta'\right)$,
$c_{mix}\left(\beta\right)$ and $c_{aux}\left(\beta'\right)$, respectively{]}
because in $\tau_{\textrm{AS}}$ the particle-2 is basically unconstrained
while the particle-$1$ is constrained to A and thus the analysis
done in the first part of Sec. \ref{sec:ClusterinEdgeWedge} can be
used. Utilizing an approach similar to that adopted in Ref. \cite{Urrutia_2013_prep}
to obtain Eq. (\ref{eq:Taui}), it is found $\tau_{\textrm{AS}}=2b_{2}V_{\textrm{S}}\frac{\beta}{\pi}-a_{2}A+2c_{aux}\left(\beta\right)L$
with $c_{aux}\left(\beta\right)=\frac{1}{2}\int_{III}g_{1}\left(z\right)d\mathbf{r}^{(2)}$
{[}see Eq. (\ref{eq:c2eHS1}){]}. Therefore, $c_{aux}\left(\beta\right)=\frac{\pi}{30}\cot\beta$
and $c_{mix}\left(\beta\right)=c_{2}\left(\beta\right)$ for all the
range $0<\beta<\pi$. Even more, $c_{2}\left(\beta'\right)=-\frac{1}{15}\left[1-\left(\pi-\beta'\right)\cot\beta'\right]$.
This is exactly the same expression given in Eq. (\ref{eq:c2eHS2}),
i.e., we have shown that Eq. (\ref{eq:c2eHS2}) applies in the extended
domain $0<\beta<\pi$.

The same approach enable us to study the case $\pi<\beta<2\pi$. In
Figure \ref{fig:edgeforAB}b it is shown the partition of the real
space U in two dihedrons with inner angles $\beta$ and $\beta'=2\pi-\beta$,
respectively. The Eqs. (\ref{eq:ProbAB}) to (\ref{eq:T2Sparted}),
that relate probabilities and cluster integrals in different regions
are still valid with the label S replaced by label U, and
\begin{equation}
\tau_{\textrm{UU}}=2b_{2}V\:,\label{eq:TauRR}
\end{equation}
where $V$ is the volume of the complete space. Thus, following the
same procedure described above one notes that Eqs. (\ref{eq:T2s-AA})
to (\ref{eq:T2Sparted2}) still apply, by changing $V_{\textrm{S}}$
by $V$ and $1/\pi$ by $1/2\pi$, while $A$ and $L$ remain unchanged.
By combining Eqs. (\ref{eq:T2s-AA2}) and (\ref{eq:T2s-BB2}) one
obtains
\begin{equation}
\tau_{\textrm{BB}}=\tau_{\textrm{BU}}+\tau_{\textrm{AA}}-\tau_{\textrm{AU}}\:.\label{eq:TauBBmAA}
\end{equation}
In this case $\tau_{\textrm{AU}}=2b_{2}V\frac{\beta}{2\pi}$ and $\tau_{\textrm{BU}}=2b_{2}V\frac{\beta'}{2\pi}$,
which implies $c_{2}\left(\beta'\right)=c_{2}\left(\beta\right)$.
Therefore,
\begin{equation}
c_{2}\left(\beta\right)=-\frac{1}{15}\left[1+\left(\pi-\beta\right)\cot\beta\right]\:,\label{eq:c2eBet}
\end{equation}
is finally valid for all the range $0<\beta<2\pi$.
\begin{figure}
\begin{centering}
\includegraphics[width=6.5cm]{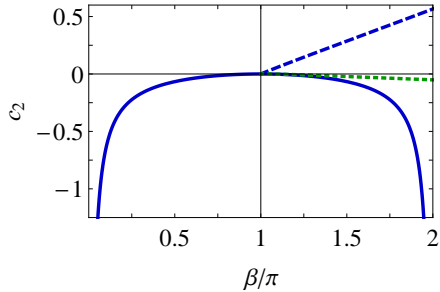}
\par\end{centering}

\protect\caption{The edge/wedge coefficient of the second cluster integral,
$c_{2}\left(\beta\right)$ for both straight- and rounded-edges. In continuous
line (blue) is plot the straight-edge result given in Eq. (\ref{eq:c2eBet}).
Other lines are for the rounded-edge case. In dashed line (blue) is plot
$c_{2}\left(\beta\right)$ from Eq. (\ref{eq:c2Fit}) (curvature correction,
d1-RR) while in dotted line (green) is plot $c_{2}\left(\beta\right)$
from Eq. (\ref{eq:c2Fitd2}) (curvature correction, d2-RR).\label{fig:c2}}
\end{figure}
In Fig. \ref{fig:c2}, it is shown $c_{2}\left(\beta\right)$ for
the straight- and rounded-edge, confinements. The expression for the
straight-edge confinement, taken from Eq. (\ref{eq:c2eBet}), is plotted
in continuous line. One observes that this $c_{2}\left(\beta\right)$
is non-positive, is symmetric around $\beta=\pi$, has smooth (analytic)
behavior at $\beta\approx\pi$ where $c_{2}$ is zero, and diverges
for both $\beta\rightarrow0$ and $\beta\rightarrow2\pi$. The results
obtained in Sec. \ref{sub:CurvatureCorrection} for $c_{2}(\beta)$
of the rounded-edge confinement are also included in Fig. \ref{fig:c2}.
There, two different flavors of the density based reference region
are presented. In dashed line it is shown d1-RR while dotted line
refers to d2-RR {[}Eqs. (\ref{eq:c2Fit}) and (\ref{eq:c2Fitd2}),
respectively{]}. All the curves reach the zero value at vanishing
edge/wedge with $\beta=\pi$.

\subsection{The rounded-edge\label{sub:CurvatureCorrection}}

Here, I analyze the effect on the HS fluid produced by a solid dihedron
that repel the core of each particle. Fig. \ref{fig:HWallEdge} shows
the edge of a solid dihedron with $\beta<\pi$ and with $\beta>\pi$
(Figs. \ref{fig:HWallEdge}a and \ref{fig:HWallEdge}b, respectively).
There, the dark region constitutes the solid wall, in white is the
excluded region induced on the HS fluid by the wall (forbidden for
the particles center), and in lighter gray (green) is the available
region for the center of particles, $\mathcal{A}$. In Fig. \ref{fig:HWallEdge}a
one can observe that the dihedral walls induce near the edge a fluid-filled
region $\mathcal{A}$ with dihedral shape, and therefore, the $c_{2}$
obtained above in Sec. \ref{sub:OtherBetaRange} applies. On the contrary,
in Fig. \ref{fig:HWallEdge}b the solid dihedron induces a curved
end-of-fluid interface region with cylindrical shape and radius $1/2$.
For this type of cylindrical-shape boundary of $\mathcal{A}$ it is
necessary to analyze the decomposition of the cluster integrals. Following
the procedure described in Ref.\cite{Urrutia_2013_prep} and focusing
on the HS fluid, in a first step one separates the integral domain
$\mathcal{A}$ over bulk and skin regions (this last in the neighborhood
of $\partial\mathcal{A}$). On a second step the planar faces domains
are separated from that in the near-edge region of $\mathcal{A}$.
Note that under this approach the surface area $A$ in Eqs. (\ref{eq:Taui},
\ref{eq:Omega}) and (\ref{eq:Nz}) is the area of the planar region
of $\partial\mathcal{A}$.

\begin{figure}
\begin{centering}
\includegraphics[clip,width=6cm]{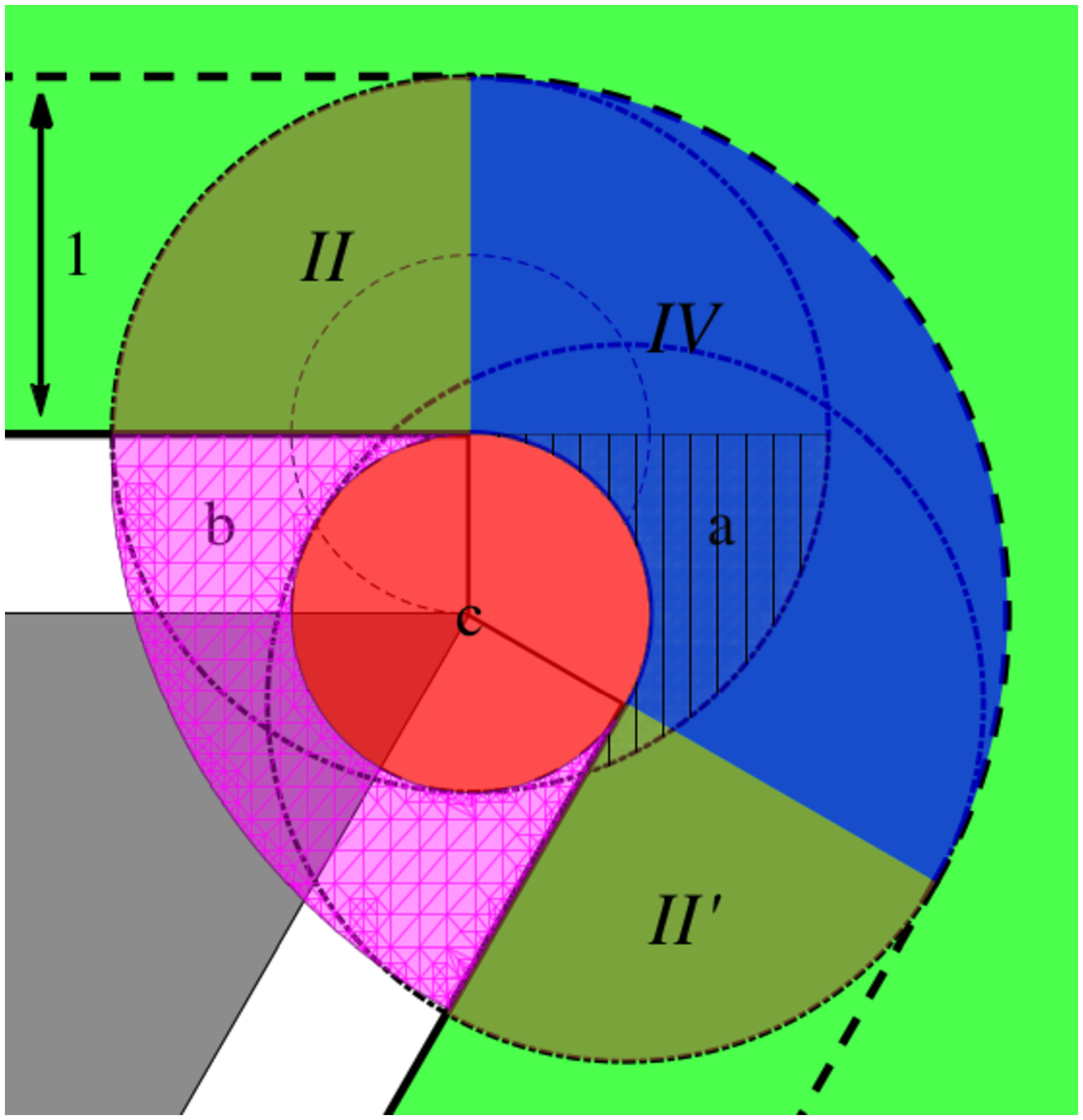}
\par\end{centering}

\protect\caption{Details of the integration domains for a rounded-edge confinement,
with $\beta>\pi$. Roman numbers ($II$, $II'$ and $IV$) label the
domain for $\mathbf{r}^{(2)}$ and $\mathbf{r}_{1}^{(2)}$ variables
while small letters label the domain for $\mathbf{r}_{2}^{(3)}$.\label{fig:EdgeCurv}}
\end{figure}

For the case of a rounded-edge confinement depicted in Fig. \ref{fig:HWallEdge}b
the overall procedure leave us again with the expression given in
Eq. (\ref{eq:c2eHS}), but now, $g_{2}$ is the volume of that part
of the exclusion sphere outside of $\mathcal{A}$ when the intersection
of the sphere with $\partial\mathcal{A}$ is non-planar. Moreover,
the domains are given in terms of regions $II$, $II'$ and $IV$
by
\begin{eqnarray}
D_{1} & = & II\:,\label{eq:D1-1}\\
D_{2} & = & II\cup II'\cup IV\:.\label{eq:D2-1}
\end{eqnarray}
These regions are shown in Fig. \ref{fig:EdgeCurv}.
Unfortunately, the involved integrals are not analytically solvable.
I found convenient to write 
\[
2c_{2}\!\left(\beta\right)=-2\int_{II}\tilde{g}_{2}\left(x\right)d\mathbf{r}^{(2)}+\int_{IV}g_{2}\left(\mathbf{r}^{(2)}\right)d\mathbf{r}^{(2)}\:,
\]
where $\tilde{g}_{2}\left(\mathbf{r}\right)$ is the volume of that
part of the exclusion sphere (centered at $\mathbf{r}$) that lies
in the region $\textrm{a}$ (shaded zone in Fig. \ref{fig:EdgeCurv}).
These integrals may be expanded as double integrals to obtain
\begin{equation}
c_{2}\!\left(\beta\right)=G+Q\:,\label{eq:c2curv}
\end{equation}
with
\begin{eqnarray}
G & = & \frac{1}{2}\int_{IV}\int_{\textrm{c}}H\, d\mathbf{r}_{2}^{(3)}d\mathbf{r}_{1}^{(2)}\:,\nonumber \\
Q & = & \frac{1}{2}\int_{IV}\int_{\textrm{b}}H\, d\mathbf{r}_{2}^{(3)}d\mathbf{r}_{1}^{(2)}-\int_{II}\int_{\textrm{a}}H\, d\mathbf{r}_{2}^{(3)}d\mathbf{r}_{1}^{(2)}\:,\label{eq:c2curv-1}
\end{eqnarray}
where $H=\Theta\bigl(1-\bigl|\mathbf{r}_{1}^{(2)}-\mathbf{r}_{2}^{(3)}\bigr|\bigr)$.
The domains b and c are also drawn in Fig. \ref{fig:IntegDom}. Finally,
one solves $c_{2}$, $G$, and $Q$ for several values of $\beta$
by MonteCarlo numerical integration.\cite{Press2007} 
\begin{figure}
\begin{centering}
\includegraphics[clip,width=6cm]{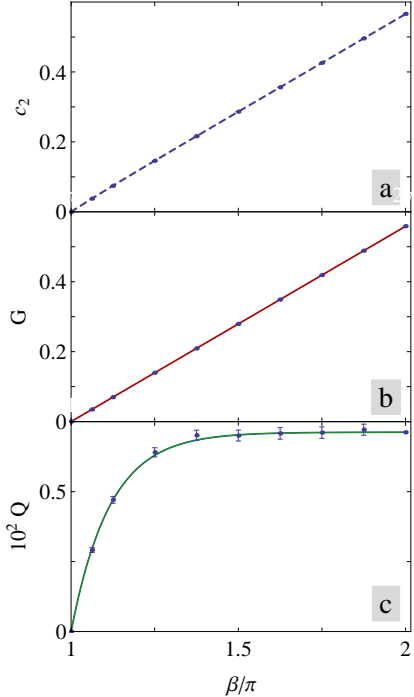}
\par\end{centering}

\protect\caption{Obtained results for $c_{2}$, $G$ and $Q$. Error bars were drawn
in each plot but in some cases are smaller than the point size. The
meaning of each line is explained in the text.\label{fig:plots}}
\end{figure}
In Fig. \ref{fig:plots} (parts a, b and c) these results are shown.
In Fig. \ref{fig:plots}a and b the nearly-linear behavior of the
points of $c_{2}\!\left(\beta\right)$ and $G$, respectively, is
apparent. To evaluate the slope of $G$ one takes circular coordinates
$d\mathbf{r}_{1}^{(2)}=r_{1}dr_{1}d\theta$ and integrates numerically
over $d\mathbf{r}_{2}^{(3)}r_{1}dr_{1}$ at arbitrary fixed angle
$\theta$. The result is $G\simeq0.17777(3)\left(\beta-\pi\right)$.
This slope coincides with the exact result $\frac{\pi}{16}*{}_{2}F_{1}\!\left(-\frac{1}{2},\frac{1}{2};3;1\right)=\frac{8}{45}$
taken from the dependence of $\tau_{2}$ on the area of the lateral
surface of a cylinder with radius $\frac{1}{2}$ {[}from Eq. (27)
in Ref. \cite{Urrutia_2010b}%
\footnote{I have detected a misprint in Eq. (27) of Ref. \cite{Urrutia_2010b}.
There, where it reads $1-{}_{2}F_{1}\!\left(-0.5,0.5;3;s^{2}\right)$
it should be $_{2}F_{1}\!\left(-0.5,0.5;3;s^{2}\right)$.%
}{]}, where $_{2}F_{1}$ is the Hypergeometric function (sometimes
denoted $F$). In Fig. \ref{fig:plots}b the \emph{exact} dependence
$G=\frac{8}{45}\left(\beta-\pi\right)$ is also plotted. A small deviation
of $c_{2}\!\left(\beta\right)$ from the linear behavior is caused
by $Q$, which is shown in Fig. \ref{fig:plots}c with the abscissa
axes augmented in a factor $10^{2}$. I fit the points using an exponential
form with two adjusting parameters and find $Q=0.007125\left\{ 1-\exp\left[-2.74\left(\beta-\pi\right)\right]\right\} $,
which is included in Fig. \ref{fig:plots}c. Therefore, the result
for $\beta>\pi$ is
\begin{equation}
c_{2}\!\left(\beta\right)=\frac{8}{45}\left(\beta-\pi\right)+Q\:,\label{eq:c2Fit}
\end{equation}
which is shown in dashed-line in Fig. \ref{fig:plots}a. This expression
describes with high precision the results of numerical integration.
For comparison, Eq. (\ref{eq:c2Fit}) is also plotted in Fig. \ref{fig:c2}
using a dashed line. The two main differences with respect to Eq.
(\ref{eq:c2eBet}) are: its positiveness and its finite value attained
in the limit $\beta\rightarrow2\pi$. This approach, that focus in
the surface area of the planar part of $\partial\mathcal{A}$, is
appointed as d1-RR.

To further analyze the obtained $c_{2}\left(\beta\right)$ in d1-RR
one can write the edge term of $\tau_{2}/2$ emphasizing its dependence
on the area of the cylindrical surface $A_{\textrm{cyl}}=\left(\beta-\pi\right)\frac{1}{2}L$.
I found
\begin{equation}
c_{2}\!\left(\beta\right)\, L=-a_{2,\textrm{cyl}}A_{\textrm{cyl}}+Q\, L\:,\label{eq:c2Fit-b}
\end{equation}
with the coefficient for the cylindrical surface area, $a_{2,\textrm{cyl}}=-\frac{16}{45}=-0.3555$.
One observes that the behavior of $c_{2}\!\left(\beta\right)\times L$
is driven by its linearity with the surface area, which resembles
the area dependence of $\tau_{2}$ and has the same sign. Alternatively,
Eq. (\ref{eq:c2Fit-b}) suggest that one may adopt a slightly different
RR by focusing on the total surface area of $\partial\mathcal{A}$.
Once the contribution $-a_{2}A_{\textrm{cyl}}$ is added (and subtracted)
to $\tau_{2}$ one obtains
\begin{equation}
c_{2}\!\left(\beta\right)=\left(\frac{8}{45}-\frac{\pi}{16}\right)\left(\beta-\pi\right)+Q\:,\label{eq:c2Fitd2}
\end{equation}
that is designated by d2-RR. Note that under this approach the surface
area $A$ in Eqs. (\ref{eq:Taui}, \ref{eq:Omega}, \ref{eq:Nz})
is the total area of $\partial\mathcal{A}$ that complains both, the
planar and cylindrical surfaces. It is interesting to highlight that
Eq. (\ref{eq:c2Fit}) and Eq. (\ref{eq:c2Fitd2}) describe the rounded-edge
confinement shown in Fig. \ref{fig:HWallEdge}b. The difference lies
in the adopted dissection of $\partial\mathcal{A}$.

\section{Results\label{sec:Results}}

Turning to the thermodynamic properties of the HS system confined
by a single edge/wedge, they can be expanded as power series in $z$
and then as power series in $\rho$. From here on, I truncate all
the series to second order and thus terms of order $O(z^{3})$ and
$O(\rho^{3})$ are depreciated. Taking into account the known values
of $b_{2}$ and $a_{2}$, one finds for the low density regime: 
\[
P/kT=\rho+\rho^{2}2\pi/3\:,
\]
\begin{eqnarray}
\gamma/kT=-\rho^{2}\pi/8 & \,\textrm{ and }\, & \Gamma_{A}=\rho^{2}\pi/4\:.\label{eq:SurfTensandAds}
\end{eqnarray}
These expressions correspond to the known expansions, of the pressure
of a \emph{bulk} HS fluid, of the fluid-wall surface tension, and
of the fluid-wall surface adsorption (when a HS fluid is confined
by a \emph{planar hard-wall}), respectively. Furthermore, the correspondence
between the obtained series for $P$, $\gamma$ and $\Gamma_{A}$,
and the known series expansions also apply when higher order terms
in $z$ and $\rho$ are considered. Therefore, Eq. (\ref{eq:SurfTensandAds})
helps to define accurately the meaning of the intensive magnitudes
introduced in Eqs. (\ref{eq:Omega}) and (\ref{eq:N}). The line-thermodynamic
properties of the HS fluid confined by an edge/wedge are
\begin{eqnarray}
\mathcal{T}/kT=-c_{2}\rho^{2} & \,\textrm{ and }\, & \Gamma_{L}=2c_{2}\rho^{2}\:.\label{eq:LineTensandAds}
\end{eqnarray}

From simple inspection of Eqs. (\ref{eq:SurfTensandAds}) and Eq.
(\ref{eq:LineTensandAds}), one notes that $\gamma/kT$ ($\mathcal{T}/kT$)
is minus one half of $\Gamma_{A}$ ($\Gamma_{L}$). It is interesting
to note that using the Gibbs adsorption equations {[}see Eq. (\ref{eq:GibbsAdsL01}){]}
up to terms of order $z^{2}$ one finds
\begin{eqnarray}
\gamma/kT=-\Gamma_{A}/2 & \,\textrm{ and }\, & \mathcal{T}/kT=-\Gamma_{L}/2\:.\label{eq:GibbsAdsL02}
\end{eqnarray}
Therefore, these remarkable relations (that holds up to terms of order
$\rho^{2}$) does not only apply to the HS fluid, but also, to \emph{any
fluid} confined by hard-wall edges and wedges. In fact, as long as
$\tau_{i}$ were given by Eq. (\ref{eq:Taui}), the relations in Eq.
(\ref{eq:GibbsAdsL02}) should be also valid for planar walls and
wedges that interact with the particles trow a less trivial external
potential $\phi\left(\mathbf{r}\right)$ than the infinite repulsion.
The adsorption equations given in Eq. (\ref{eq:GibbsAdsL02}) are
very similar to the ideal gas equation of state $P/kT=\rho$ in several
aspects. In particular, they are simple relations between free-energy
densities, number densities and temperature, i.e. they are simple
\emph{equations of state} of the adsorbed gas. These ideal-adsorption
equations include terms up to order $\rho^{2}$, terms of the same
order are not included in the pressure of the ideal gas.

\begin{figure}
\begin{centering}
\includegraphics[width=7.5cm]{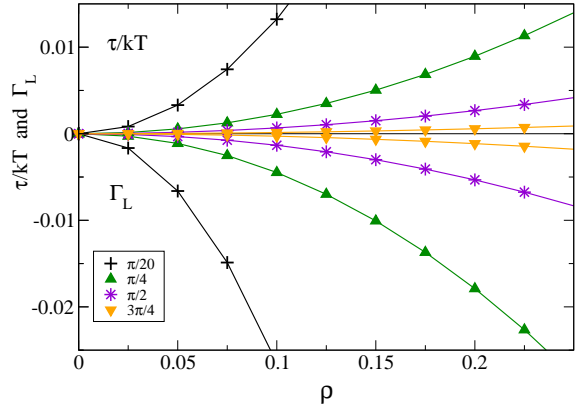}
\par\end{centering}

\protect\caption{Line adsorption and line tension (scaled with temperature) for the
straight-edge confinement at four angles in the range $0<\beta<\pi$.
Line adsorption curves lie in the negative abscissa region, while
line tension curves lie in the positive ones.\label{fig:Line-adsandtens}}
\end{figure}
 In Fig. \ref{fig:Line-adsandtens} the density dependence of line
tension (scaled with temperature) and line-adsorption for several
edges/wedges with $\beta<\pi$, are shown (see Figs. \ref{fig:Adihedron}a
and \ref{fig:HWallEdge}a). Different curves correspond to four different
angles. There, the line tension curves are positive but line adsorption
ones are negatives. The curves for larger values of $\beta$ are near
to the zero ordinate axis, which corresponds to the limiting case
$\beta=\pi$ of an edge that vanishes in an infinite planar wall.
On the contrary, in the limit $\beta\rightarrow0$ both $\mathcal{T}/kT$
and $\Gamma_{L}$ diverge (not shown in the Figure) because $c_{2}(\beta)$
diverges. This can be rationalized as the impossibility of attaining
the dimensional crossover to the hard disc system with Eq. (\ref{eq:c2Fit})
because an extra non-analytic term should appears in the case of $\beta\rightarrow0$.
This was previously verified for the HS system under confinement in
box-shaped cavities.\cite{Urrutia_2010b} Note that, no matter the
value of $\beta$ (in the range $0<\beta<\pi$) the HS system \emph{apparently
desorbs} from the edge/wedge region (it has a negative adsorption).
This unexpected behavior comes from the direct analysis of the pure
line-adsorption term $\Gamma_{L}$ given in Eq. (\ref{eq:N}). I will
return to this point at the end of this section.

\begin{figure}
\begin{centering}
\includegraphics[width=7.5cm]{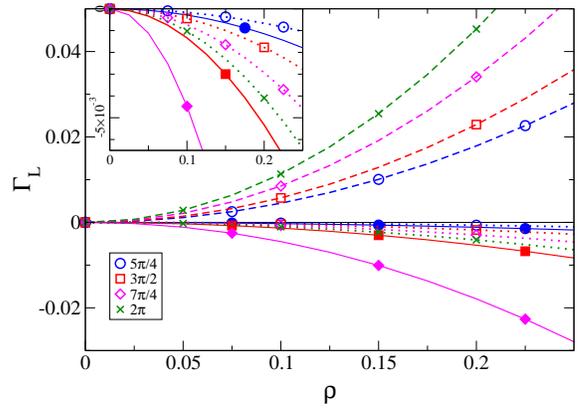}
\par\end{centering}

\protect\caption{Line adsorption for both, the straight- and rounded-edges confinement,
at angles in the range $\pi<\beta\leq2\pi$. Continuous lines (with
filled symbols) show results for the straight-edge confinement while
dashed and dotted lines correspond to rounded-edge (d1-RR and d2-RR,
respectively).\label{fig:Line-adsorption}}
\end{figure}
Fig. \ref{fig:Line-adsorption} presents the behavior of line-adsorption
with density for different edges/wedges that correspond to $\beta>\pi$.
In the inset, the details of a region near the horizontal axis is
shown. Straight lines correspond to the straight-edge confinement
(see Fig. \ref{fig:Adihedron}b) while discontinuous lines correspond
to the rounded-edge case (see Fig. \ref{fig:HWallEdge}b). The dashed
and dotted lines correspond to the different criteria adopted for
the dissection of $\partial\mathcal{A}$, d1-RR and d2-RR, respectively.
I wish to note some interesting features of Fig. \ref{fig:Line-adsorption}
that follows from the obtained $c_{2}\left(\beta\right)$ results
(see Fig. \ref{fig:c2}). First, one observes that for a fixed $\beta$
(and considering d1-RR), the pure line-adsorption on straight- and
rounded-edges are very different (they have opposite sign). Second,
for the rounded-edge the line-adsorption obtained using d1-RR differs
strongly from that obtained by using d2-RR. Thus, based on these issues
one concludes that straight-edges behaves very different to rounded-edges,
and also, that given a fixed physical constraint the adoption of different
RRs may produce strongly different results. A third noticeable characteristic
is that straight-edges show \emph{apparent desorption} for any (nontrivial)
dihedral angle. A comparison between Fig. \ref{fig:Line-adsandtens}
and Fig. \ref{fig:Line-adsorption} shows that this \emph{apparent
desorption is symmetric} around $\beta=\pi$, i.e. $\Gamma_{L}(\pi-\Delta\beta)=\Gamma_{L}(\pi+\Delta\beta)$.

Turning to Fig. \ref{fig:Line-adsorption}, one notes that in all
cases the curves for smaller values of $\beta$ are nearer to the
zero ordinate axis (which gives the limiting planar-wall case $\beta=\pi$
of a vanishing edge). The other interesting limit for $\Gamma_{L}$
is $\beta\rightarrow2\pi$. There, the curve for the rounded-edge
does not diverge, but $\Gamma_{L}$ for the straight-edge diverges
(not shown in the figure) because $c_{2}(\beta)$ {[}see Eq. (\ref{eq:c2eHS}){]}
diverges too. This can be rationalized again by recognizing that Eq.
(\ref{eq:c2Fit}) is unsuitable to analyze the confinement produced
by a vanishing straight-wedge, because an extra non-analytic term
should appear in the case of $\beta\rightarrow2\pi$. In this limit
of the straight-edge case, the available region for the fluid $\mathcal{A}$
becomes the complete space with volume $V$ (in fact one should subtract
half plane, that has zero volume measure), and thus, the homogeneous
system result $\tau_{2}=2b_{2}V$ should be recovered.
\begin{figure}
\begin{centering}
\includegraphics[width=7.5cm]{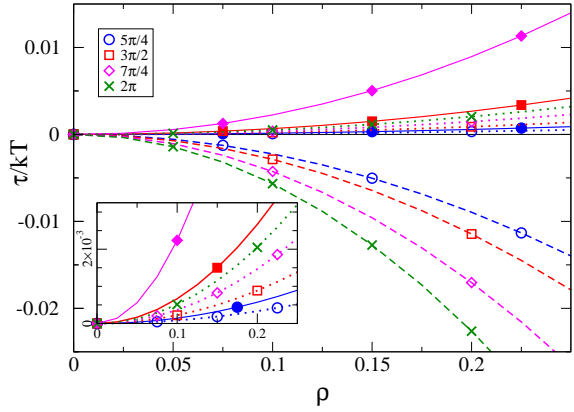}
\par\end{centering}

\protect\caption{Line tension (scaled with temperature) for both, the straight- and
rounded-edges confinement for four angles in the range $\pi<\beta\leq2\pi$.
See details in Fig. \ref{fig:Line-adsorption}.\label{fig:Line-tension}}
\end{figure}
 In Fig. \ref{fig:Line-tension} the dependence of line-tension with
density for different edges/wedges with $\pi<\beta\leq2\pi$, is shown.
The inset shows a detail of a region near the axis. One notes that
for the straight-edge confinement the line tension (i.e. the line
component of the Grand free-energy) monotonically increases with increasing
$\beta$ (from $\beta=\pi$) and diverges for all densities at $\beta\rightarrow2\pi$
because this limit can not be described by Eq. (\ref{eq:c2eHS}).
The reasons of this behavior are the same discussed in the case of
Fig. \ref{fig:Line-adsorption}. Other features observed in Fig. \ref{fig:Line-tension}
follow the characteristics noted for the curves in Fig. \ref{fig:Line-adsorption}.
Perhaps, the most interesting result about $\mathcal{T}$ is not in
the plotted curves in Fig. \ref{fig:Line-tension} but on the existence
of the analytic expression for $\mathcal{T}$, in fact, it enable
us to give the analytic expression of the grand free energy $\Omega\left(\beta,\rho\right)$
up to order $\rho^{2}$.

The unexpected \emph{apparent desorption} and \emph{symmetry} of $\Gamma_{L}$
for system confined by a straight-edge, and the strong dependence
of $\Gamma_{L}$ on the adopted RR for system confined by a rounded-edge,
both suggest an alternative approach to analyze the adsorption produced
by edges and wedges. I realize that a more meaningful definition of
edge/wedge adsorption that considers finite volumes and sets as reference
the planar wall is possible. Thus, I introduce the mean excess of
adsorbed density 
\begin{eqnarray}
\Delta\rho & = & N(r)/V(r)-\rho\:.\label{eq:Dltrho}
\end{eqnarray}
where $N(r)$ is the mean number of particles in a suitably selected
region with volume $V(r)$ and thickness $r$. A similar approach
enable us to introduce the mean excess of free-energy density
\begin{eqnarray}
\Delta\omega & = & \Omega(r)/V(r)+P\:,\label{eq:Deltaomega}
\end{eqnarray}
here $\Omega(r)$ is the free-energy of the $N\left(r\right)$ particles
in $V\left(r\right)$. Both $\Delta\rho$ and $\Delta\omega$ quantify
the local behavior of the HS system in the neighborhood of the apex.
The analytic expressions of the mean excess densities $\Delta\rho$
and $\Delta\omega$ for the straight-edge and for the rounded-edge
are deduced in the Appendix \ref{Appdx:Geometric}.
\begin{figure}
\centering{}\includegraphics[width=7.5cm]{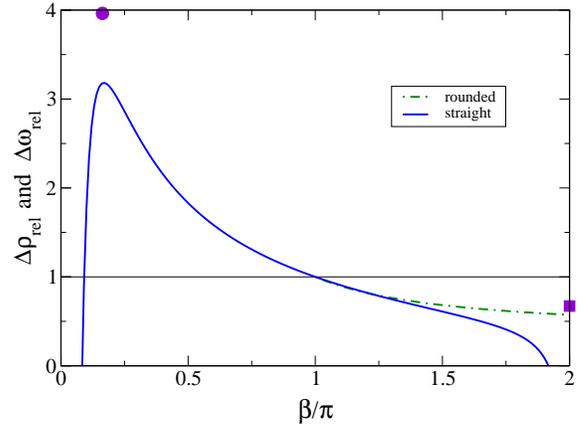}\protect\caption{Relative mean excess adsorbed density and free-energy density in the
neighborhood of the edge/wedge divided by its planar wall value. $\Delta\rho_{rel}=\Delta\omega_{rel}$
and they are independent of the density up to the considered order.\label{fig:EdgeMdensity}}
\end{figure}
Both, $\Delta\rho$ and $\Delta\omega$ are proportional to $\rho^{2}$
(terms of higher order are not considered in current work). Under
the adopted truncation of terms $O(\rho^{3})$ the distance between
$\partial\mathcal{A}$ and the zone of the spatial density distribution
where it attains its bulk value $\rho$ is $1$, thus, I fixed $r=1$.
It is convenient to introduce the relative mean excess values of the
adsorbed density and free-energy density, 
\begin{equation}
\Delta\rho_{rel}=\frac{\Delta\rho}{\Delta\rho_{wall}}\:\textrm{and}\:\Delta\omega_{rel}=\frac{\Delta\omega}{\Delta\omega_{wall}}\:,\label{eq:DrhoDomg}
\end{equation}
respectively. One can note some properties of $\Delta\rho_{rel}$
and $\Delta\omega_{rel}$: they are independent of the number density
$\rho$ and only depend on $\beta$, their analytic expressions are
the same i.e. $\Delta\rho_{rel}=\Delta\omega_{rel}$, and if the d2-RR
is adopted (in place of d1-RR) the expressions for $\Delta\rho_{rel}$
and $\Delta\omega_{rel}$ remain unmodified. An important difference
between $\Delta\rho_{rel}$ and $\Delta\omega_{rel}$ is the sign
of the denominator in Eq. (\ref{eq:DrhoDomg}). Given that $\Delta\rho_{wall}>0$
it follows that $\Delta\rho_{rel}>0\Leftrightarrow\Delta\rho>0$.
On the contrary, $\Delta\omega_{wall}<0$ and thus it follows that
$\Delta\omega_{rel}>0\Leftrightarrow\Delta\omega<0$. This implies
that a maximum in $\Delta\omega_{rel}$ indicates a minimum in $\Delta\omega$.
In Fig. \ref{fig:EdgeMdensity} the relative magnitudes $\Delta\rho_{rel}$
and $\Delta\omega_{rel}$ are plotted. Continuous line shows the result
for the straight-edge confinement {[}see Eq. (\ref{apeq:Drhorel}){]}
while the dot-dashed line corresponds to the rounded-edge. Note that,
the symmetry found in $c_{2}(\beta)$, $\Gamma_{L}$ and $\mathcal{T}/kT$
for the straight-edge {[}$c_{2}(\beta)=c_{2}(2\pi-\beta)$ for $0<\beta<\pi${]}
disappears in $\Delta\rho_{rel}$ and $\Delta\omega_{rel}$. The curves
show that in the neighborhood of a very acute edge ($0<\beta\lesssim0.091\pi$)
the HS gas desorbs, as it also happens for the gas near a solid wedge
($\beta>\pi$). In the intermediate range $0.091\pi\lesssim\beta<\pi$
that includes acute and obtuse edges the HS gas adsorbs, with a peak
of maximum adsorption at $\beta\approx0.17\pi$ with $\Delta\rho_{rel}\approx3.2$.
All these trends agree with that found in Ref. \cite{Schoen_1997},
where the single bulk density $\rho=0.7016$ (which is not a small
density because fluid-solid transition is at $\rho\approx1$) was
studied using free-energy density functional theory (DFT). Besides,
the local mean density of free-energy relative to the planar-wall
condition follows exactly the same behavior that shows $\Delta\rho_{rel}$,
which implies a minimum of mean excess of free-energy density $\Delta\omega$
at $\beta\approx0.17\pi$. The results presented in Fig. \ref{fig:EdgeMdensity}
slightly depend on the details of the adopted region $V(r)$ and on
the adopted value of $r$. Although, the discussed general trends
remain unmodified. As an indication of this feature, the values obtained
using a different definition of $V\left(r\right)$ are shown with
a small circle (the maximum) and a small square (the rounded-edge
value for $\beta=2\pi$). More details are given in Appendix \ref{Appdx:Geometric}.

\subsection{Adsorbed HS gas on a Corrugated wall}

The results shown in Fig. \ref{fig:EdgeMdensity} motivated the study
of a more complex geometrical confinement than edges and wedges. Here,
I analyze the properties of the HS gas on contact with a corrugated
wall in the low density regime, by focusing in the total adsorption
and surface free-energy of the system. Adsorption is an easy measurable
magnitude accessible through experiments, through a variety of simulations
techniques (molecular dynamics and MonteCarlo) and also using DFT.
On the other hand, the surface free-energy density is not simple to
measure in experiments but it can be approximately evaluated by DFT.
Together, they provide a variety of quantitative and high-precision
tests between different approaches. 
\begin{figure}
\begin{centering}
\includegraphics[width=5.5cm]{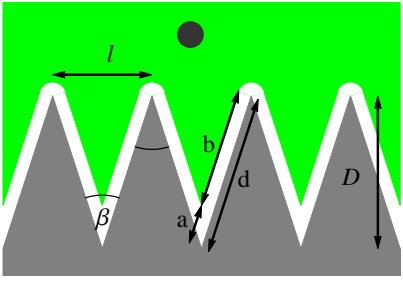}
\par\end{centering}

\protect\caption{Scheme of the corrugated wall built by acute straight-edges (grooves)
and rounded-edges (ridges).\label{fig:CorrugatedWall}}
\end{figure}
 In Fig. \ref{fig:CorrugatedWall} a draw of the system with its principal
lengths is shown and a single HS particle is included. There, $l$
is the distance between cups and $D$ is the depth. Lengths $a$,
$b$, and $d$ are used to calculate the relevant areas in Appendix
\ref{Appdx:DetailedGeom}. This complex-shape confinement includes
rounded-edges as that shown in Fig. \ref{fig:HWallEdge}b. In the
analysis it is adopted the d1-RR point of view. Given a planar wall
of area $A_{0}$, the number of adsorbed particles per unit area is
simply $N_{A}/A_{0}=\Gamma_{A}$. For a corrugated wall scratched
on the planar wall one defines the effective adsorption
\begin{equation}
\Gamma_{eff}=\frac{N_{A}}{A_{0}}=\Gamma_{A}\frac{A}{A_{0}}+\left(\Gamma_{L}+\Gamma_{L}^{'}\right)\frac{L_{tot}}{A_{0}}\:,\label{eq:Gameff}
\end{equation}
with $\Gamma_{L}$ the line adsorption of the groove edge ($\beta<\pi$),
$\Gamma_{L}^{'}=\Gamma_{L}(2\pi-\beta)$ the line adsorption of the
ridge edge, and $L_{tot}$ the total length of the edges with dihedral
angle $\beta$. For the analyzed HS system the magnitudes $\Gamma_{A}$
and $\Gamma_{L}$ were given in Eqs. (\ref{eq:SurfTensandAds}, \ref{eq:LineTensandAds}),
while the analytic expressions of $A$ and $A_{0}$ as functions of
$\beta$ and $D$ are evaluated in the Appendix \ref{Appdx:DetailedGeom}.
Following the same approach one defines the effective surface density
of free-energy
\begin{equation}
\gamma_{eff}=\frac{\Omega_{A}}{A_{0}}=\gamma\frac{A}{A_{0}}+\left(\mathcal{T}+\mathcal{T}^{'}\right)\frac{L}{A_{0}}\:,\label{eq:gameff}
\end{equation}
with $\Omega_{A}=\Omega+PV$ and $\mathcal{T}^{'}=\mathcal{T}\left(2\pi-\beta\right)$.
Both $\Gamma_{eff}$ and $\gamma_{eff}$ are proportional to $\rho^{2}$
and depend on a non-trivial way on $\beta$ and $D$. On the other
hand, the relative adsorption and the relative surface density of
free-energy 
\begin{equation}
\Gamma_{rel}=\frac{\Gamma_{eff}}{\Gamma_{A}}\:\textrm{and}\:\gamma_{rel}=\frac{\gamma_{eff}}{\gamma}\:,\label{eq:Gamgamrel}
\end{equation}
respectively, are both independent of $\rho$. Also, they are identical
i.e. $\Gamma_{rel}=\gamma_{rel}$.
\begin{figure}
\begin{centering}
\includegraphics[width=7.5cm]{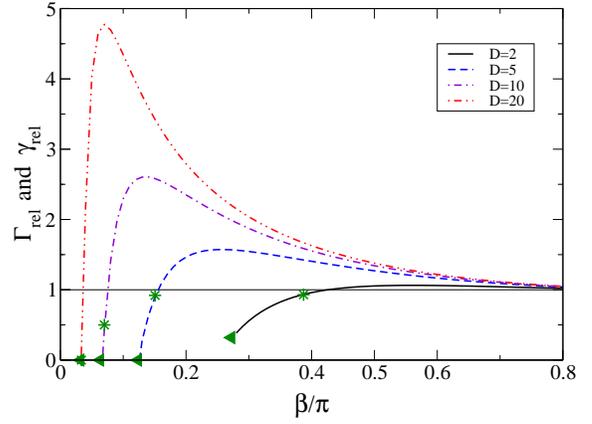}
\par\end{centering}

\protect\caption{Relative adsorption and relative free-energy per surface area for
the corrugated wall as a function of dihedral angle of the groove
edge. $\Gamma_{rel}=\gamma_{rel}$ and they are independent of the
density up to the considered order. Curves for several values of the
depth $D$ are shown.\label{fig:relAdsndOmg}}
\end{figure}
 In Fig. \ref{fig:relAdsndOmg} $\Gamma_{rel}$ and $\gamma_{rel}$
are presented as functions of $\beta$ and $D$. In the limit of vanishing
edge ($\beta\rightarrow\pi$) they go, naturally, to one. On the contrary,
for very small dihedral angles curves show that the system desorbs
from the corrugated wall ($\Gamma_{rel}<1$) and consistently the
relative surface density of free energy also becomes smaller than
one. In the intermediate angular range the relative adsorption (and
the relative free energy) has a peak which becomes more and more pronounced
and goes to smaller values of $\beta$ as $D$ increase. A cutoff
on the confidence of the plotted curves is introduced through the
use of stars and triangles. Stars show the smaller value of $\beta$
(for each $D$) beyond that the ridge and the groove become so nearer
that their interference could be noted (it was estimated using the
condition $b>1$). This interference becomes relevant for corrugated
walls with short spatial periodicity length $l$. In such case an
extra interference term $\Delta c_{2}$ that couples ridge and groove
contributions should be considered. This term should compensate the
divergence of $c_{2}\left(\beta\right)$ for very small angles ($\beta\rightarrow0$).
Plotted triangles mark the smaller value of $\beta$ (for each $D$)
beyond that one expects that $\Delta c_{2}$ becomes the relevant
term (it was estimated using the condition $b>0$). The evaluation
of $\Delta c_{2}$ is beyond the scope of present work.

\section{Final Remarks\label{sec:End}}

In this work the thermodynamic properties of the confined HS fluid,
which is a relevant reference system for both simple and colloidal
fluids, was studied. The HS fluid was confined in wedges and by edges,
and analyzed in the framework of the activity series expansion for
the grand free energy. The coefficients of these series are the inhomogeneous
version of the Mayer's cluster integrals, which decompose linearly
in terms of its volume, surface area, and edges length components.
Two different type of edge/wedge confinement were analyzed: the straight-edge
and the rounded-edge.

On the basis of this non-standard approach and using analytic grounds
it was studied in the low density regime the dependence of the linear-thermodynamic
properties on the dihedral angle. This analysis was done after obtaining
the functional dependence of the second cluster integral, $\tau_{2}$,
with the dihedral angle for the complete range $0<\beta<2\pi$ and
for both types of edges. In particular, the second cluster edge component
$c_{2}(\beta)$ was obtained by applying analytic and numerical integration
schemes, and a statistical-based analysis.

The exact expression for $c_{2}(\beta)$ was derived in the case where
the center of each particle lies in a straight-edge region. In addition,
accurate analytic expressions for $c_{2}(\beta)$ were found in the
case of a hard-wall wedge that induce a rounded-edge confinement ($\pi<\beta<2\pi$).
The up to present unknown $c_{2}(\beta)$ allows to obtain the pure
line-tension and line-adsorption of a single edge/wedge in terms of
density power series up to order two. The analytic approach to the
edge/wedge confined HS system was extended to evaluate the mean excess
of, adsorbed density and free-energy density in the neighborhood of
the edge, relative to the planar wall. Both properties exhibit an
interesting peak at $\beta\approx\pi/6$ that corresponds to a maximum
of adsorption and minimum of free-energy. Furthermore, formulas were
also extended to analyze the HS system confined by a corrugated wall.
In this case, it was observed an interesting non-trivial behavior
of the effective adsorption and effective surface free-energy that
develop an extremum as function of the dihedral angle and the depth
of the corrugation.

Notably, the obtained results for HS are absolute in the sense that
they do not involve any assumption about approximate equations of
state for the bulk system, and thus, they constitute reference values
for the validation and development of more accurate approximate theories
for inhomogeneous fluid. Further, the expressions found for $c_{2}$
of HS suggest that the system of square well particles may be also
studied by a direct extension of the developed scheme. Finally, concerning
the HS system, both the statistical-based and the MonteCarlo approaches
that were implemented to evaluate $c_{2}$ are promising to extend
the studies to $c_{3}$ and $c_{4}$ (third and fourth cluster integrals).
\begin{acknowledgments}
This work was supported by Argentina Grants ANPCyT PICT-2011-1887
and CONICET PIP-112-200801-00403.%

\end{acknowledgments}

\appendix

\section{Mean excess densities $\Delta\rho$ and $\Delta\omega$ for a single
edge/wedge\label{Appdx:Geometric}}

To obtain the analytic expressions of $\Delta\rho$ and $\Delta\omega$
one should choose a small and simple region around the apex. For this
region with thickness $r$ one assumes that $\varOmega\left(r\right)$
and $N\left(r\right)$ have a linear form like those given in Eqs.
(\ref{eq:Omega}, \ref{eq:N}), in terms of measures $V\left(r\right)$,
$A\left(r\right)$ and $L$. Therefore one finds
\begin{equation}
\Delta\rho=\frac{A\left(r\right)}{V\left(r\right)}\Gamma_{A}+\frac{L}{V\left(r\right)}\Gamma_{L}\:,\label{apeq:Drho}
\end{equation}
\begin{equation}
\Delta\omega=\frac{A\left(r\right)}{V\left(r\right)}\gamma+\frac{L}{V\left(r\right)}\mathcal{T}\:.\label{apeq:Dw}
\end{equation}
\begin{figure}
\begin{centering}
\includegraphics[width=7.5cm]{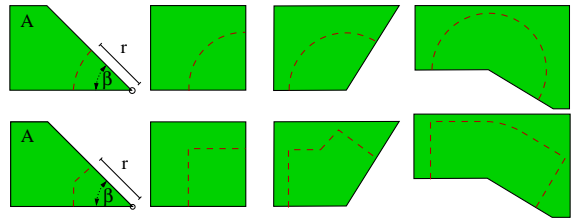}
\par\end{centering}

\protect\caption{Regions used to evaluate the mean values over finite volumes, $\Delta\rho$
and $\Delta\omega$, for different values of the dihedral angle $\beta$.\label{apfig:Geom}}
\end{figure}
For the straight-edge the adopted region is that part of a cylinder
with radius $r$ and axis on the edge that intersects $\mathcal{A}$,
as can be observed in Fig. \ref{apfig:Geom} first row. Thus, $A(r)/V(r)=\frac{4}{\beta r}$
and $L/V(r)=\frac{2}{\beta r^{2}}$. For the rounded-edge I adopt
d1-RR and take the region with volume $V(r)$ described in Fig. \ref{fig:EdgeCurv}
as region $II+II'+IV$ (with depth $r$ in place of $\sigma$). Therefore,
$V\left(r\right)=L\left[\frac{\beta-\pi}{2}\left((r+\frac{1}{2})^{2}-\frac{1}{4}\right)+\frac{\pi}{2}r^{2}\right]$
and $A\left(r\right)=L2r$. Note that the truncated low order density
expansion was not yet introduced in Eqs. (\ref{apeq:Drho}, \ref{apeq:Dw}).
Now, one replaces in Eq. (\ref{apeq:Drho}) {[}Eq. (\ref{apeq:Dw}){]}
$\Gamma_{A}$ and $\Gamma_{L}$ ($\gamma$ and $\mathcal{T}$) with
the exact expressions up to order $\rho^{2}$ taken from Eqs. (\ref{eq:SurfTensandAds},
\ref{eq:LineTensandAds}). To fix the value of $r$ one relates it
with the range of the inhomogeneous density profile consistent with
the order $\rho^{2}$ truncation, i.e. $\sigma$. In principle it
would be $r\gtrsim\sigma$ but for simplicity It is adopted $r=\sigma$.
For the straight-edge one obtains the simple analytic expression 
\begin{equation}
\Delta\rho_{rel}=\Delta\omega_{rel}=\frac{\pi}{\beta}+\frac{4}{\beta}c_{2}\:,\label{apeq:Drhorel}
\end{equation}
{[}$c_{2}$ is given in Eq. (\ref{eq:c2eHS2}){]} which is plotted
in Fig. \ref{fig:relAdsndOmg}.

The above adopted region around the apex of the edge or wedge, is
not the unique possible choice. A more subtle analysis shows that
the following more complex alternative, which modifies $V\left(r\right)$
(but not the area $A\left(r\right)=L2r$), may be better. For the
case of a straight-edge the regions are shown in Fig. \ref{apfig:Geom}
second row. One finds the volume definition: If $0<\beta<\pi/2$ then
$V(r)=Lr^{2}\tan\left(\beta/2\right)$, if $\pi/2<\beta<\pi$ then
$V(r)=Lr^{2}\left[2-\cot\left(\beta/2\right)\right]$, while if $\pi<\beta<2\pi$
then $V(r)=Lr^{2}\left[2+\beta/2\right]$. On the rounded-edge case,
one adopts a region akin that depicted in Fig. \ref{fig:EdgeCurv}
as region $II+II'+IV$ but with \foreignlanguage{english}{$II$} and
\foreignlanguage{english}{$II'$} replaced by squares with sides of
length $r$. It gives $V(r)=L\left[\frac{\beta-\pi}{2}\left((r+\frac{1}{2})^{2}-\frac{1}{4}\right)+2r^{2}\right]$.
Finally one fixes $r=\sigma$.

\section{Relative adsorption and surface free-energy on a corrugated wall\label{Appdx:DetailedGeom}}

The characteristic lengths used to develop explicit formulas for the
areas $A_{0}$ and $A$ are presented in Fig. \ref{fig:CorrugatedWall}.
The area of the planar wall surface of reference is $A_{0}=l*Ln$
where $n$ is the number of grooves separated by a distance $l$ that
will be considered and $L_{tot}=Ln$. This planar wall is transformed
to a corrugated wall by scratching. For the corrugated wall I adopt
the d1-RR {[}see Eq. (\ref{eq:c2Fit}){]} and thus the relevant area
$A$ is the area of the planar part of the gas-wall interface. In
terms of the lengths shown in Fig. \ref{fig:CorrugatedWall} it is
obtained: $A=2b*Ln$ with $b=d-a$, $a=\sigma\cot\left(\beta/2\right)$,
$d=D\sec\left(\beta/2\right)$ and $l=2D\tan\left(\beta/2\right)$.
Rearranging these identities one finds
\begin{eqnarray}
A/Ln & = & 2\left[D\sec\left(\beta/2\right)-\sigma\cot\left(\beta/2\right)\right]\:,\nonumber \\
A_{0}/Ln & = & 2D\tan\left(\beta/2\right)\:.\label{apeq:AyA0}
\end{eqnarray}
Once Eq. (\ref{apeq:AyA0}) is replaced in Eqs. (\ref{eq:Gameff},
\ref{eq:gameff}) one obtains
\begin{equation}
X_{rel}=\csc\left(\beta/2\right)-\frac{\sigma}{D}\cot\left(\beta/2\right)^{2}+y_{rel}\frac{1}{D}\cot\left(\beta/2\right)\:,\label{apeq:Xrel}
\end{equation}
where the relative magnitude $X_{rel}$ is any of $\Gamma_{rel}$
and $\gamma_{rel}$, and $y_{rel}$ is $\left(\Gamma_{L}+\Gamma_{L}^{'}\right)/\Gamma_{A}$
or $\left(\mathcal{T}_{L}+\mathcal{T}_{L}^{'}\right)/\gamma_{A}$,
in each case. Note that $\sigma$ is here related to the external
confining potential, it is the minimum enabled distance between the
center of the particle and the substrate. Naturally, for the HS fluid
$\sigma$ is also the pair collision distance. The two points of view
unify when one calls $\sigma$ as the particle diameter. Finally,
for the HS up to order $\rho^{2}$ it is $y_{rel}=8\left[c_{2}\left(\beta\right)+c_{2}\left(2\pi-\beta\right)\right]/\pi$
with $c_{2}\left(\beta\right)$ and $c_{2}\left(2\pi-\beta\right)$
taken from Eqs. (\ref{eq:c2eHS2}, \ref{eq:c2Fit}). This implies
that $X_{rel}$ becomes independent of $\rho$.

If the d2-RR is adopted in place of d1-RR, the relevant area for the
corrugated wall turns to be $A/Ln=2b+\sigma\left(\pi-\beta\right)/2$
{[}see Eq. (\ref{eq:c2Fitd2}){]}. Furthermore, $y_{rel}$ changes
because $c_{2}\left(2\pi-\beta\right)$ is that of Eq. (\ref{eq:c2Fitd2}).
These changes do not modify $X_{rel}$ and thus Fig. \ref{fig:relAdsndOmg}
also shows those results found under the adoption of d2-RR.

%%\bibliographystyle{aipnum4-1}
%%\bibliography{/home/urrutia/bibtexref/referencias_nacho_2014}

%

\end{document}